\documentclass[aps,prd,superscriptaddress,groupedaddress,nofootinbib]{revtex4}

\usepackage{graphicx}
\usepackage{dcolumn}
\usepackage{bm}
\usepackage{inputenc}
\usepackage{float}
\usepackage{amsmath}
\usepackage{amssymb}
\usepackage{mathtools}
\usepackage{siunitx}
\usepackage{amsfonts}
\usepackage{dsfont}
\usepackage{array}
\usepackage{mathrsfs}
\usepackage{pifont}
\usepackage{multirow}
\usepackage{upgreek}
\usepackage[dvipsnames]{xcolor}
\usepackage{hyperref}
\usepackage{verbatim}
\usepackage{slashed}
\usepackage{cleveref}
\usepackage{ucs}
\usepackage{tikz}
\usepackage{subfigure}
\usepackage{caption}
\usepackage{makecell}
\usepackage{rotating}
\usepackage{makecell}
\usepackage{slashed}
\usepackage{adjustbox}
\usepackage[compat=1.1.0]{tikz-feynman}
\usepackage{tikz-feynhand}

\usepackage[normalem]{ulem}

\begin{document}

\title {Collins-Soper Kernel and Reduced Soft Function  in Lattice QCD}

\author{Constantia Alexandrou}
\email{alexand@ucy.ac.cy}
\affiliation{Department of Physics, University of Cyprus, P.O. Box 20537, 1678 Nicosia, Cyprus}
\affiliation{Computation-based Science and Technology Research Center, The Cyprus Institute, 20 Kavafi Str., Nicosia 2121, Cyprus}
\author{Simone Bacchio}
\affiliation{Computation-based Science and Technology Research Center, The Cyprus Institute, 20 Kavafi Str., Nicosia 2121, Cyprus}
\author{Krzysztof Cichy}
\affiliation{Faculty of Physics and Astronomy, Adam Mickiewicz University, ul. Uniwersytetu Poznańskiego 2, 61-614 Poznań, Poland}
\author{Martha Constantinou}
\affiliation{Temple University,1925 N. 12th Street, Philadelphia, PA 19122-1801, USA}
\author{Aniket Sen}
\affiliation{Helmholtz Institut für Strahlen- und Kernphysik, Rheinische Friedrich-Wilhelms-Universität Bonn, Nussallee 14-16, 53115 Bonn}
\affiliation{Bethe Center for Theoretical Physics, Rheinische Friedrich-Wilhelms-Universität Bonn, 
Nussallee 12, 53115 Bonn, Germany}
\author{Gregoris Spanoudes}
\affiliation{Department of Physics, University of Cyprus, P.O. Box 20537, 1678 Nicosia, Cyprus}
\author{Fernanda Steffens}
\affiliation{Helmholtz Institut für Strahlen- und Kernphysik, Rheinische Friedrich-Wilhelms-Universität Bonn, Nussallee 14-16, 53115 Bonn}
\affiliation{Bethe Center for Theoretical Physics, Rheinische Friedrich-Wilhelms-Universität Bonn, 
Nussallee 12, 53115 Bonn, Germany}
\author{Jacopo Tarello}
\email{j.tarello@cyi.ac.cy}
\affiliation{Department of Physics, University of Cyprus, P.O. Box 20537, 1678 Nicosia, Cyprus}
\affiliation{Computation-based Science and Technology Research Center, The Cyprus Institute, 20 Kavafi Str., Nicosia 2121, Cyprus}

\date{\today}

  \begin{abstract} 
We evaluate the Collins-Soper kernel and the reduced soft function in lattice QCD, incorporating $\mathcal{O}(\alpha_s)$ matching corrections. The calculation relies on the evaluation of the quasi-transverse momentum-dependent wave function with asymmetric staple-shaped quark bilinear operators and four-point meson form factors. These quantities are computed non-perturbatively using two $N_f=2+1+1$ twisted-mass fermion ensembles with the same lattice spacing  of $a=0.093$ fm: the first ensemble has a lattice size of $24^3 \times 48$ and a pion mass of 346~MeV; the second one has a lattice size of $32^3 \times 64$ and a pion mass of 261~MeV.  The Collins-Soper kernel and the soft function are needed for the determination of the transverse momentum-dependent parton distribution functions.
\end{abstract}
   
\maketitle 

    \section{INTRODUCTION}
    \label{sec:introduction}
	
The description of the internal structure of
hadrons using the fundamental degrees of freedom 
of QCD has been a central focus of
nuclear physics in the last 50 years,
both theoretically and 
experimentally, particularly 
through the study of the so-called
parton distribution functions (PDFs). 
PDFs are one-dimensional objects describing
how the partons are distributed relative to the
fraction of the longitudinal momentum of the parent
hadron that they 
carry.
To go beyond this one-dimensional picture, one
needs to compute generalized parton distributions or take into account the transverse motion
of partons within a hadron, and this information is
encapsulated in the transverse momentum-dependent parton
distribution functions (TMD~PDFs)~\cite{Angeles-Martinez:2015sea,Boussarie:2023izj}.
In this work, we will be addressing TMD~PDFs.
Experimentally, TMD~PDFs can be
accessed via Drell-Yan scattering~\cite{Collins:1981uk,Collins:1981va,Collins:1984kg,Collins:2017oxh} or via
semi-inclusive deep-inelastic scattering (SIDIS)~\cite{Ji:2004wu,Ji:2004xq}, and their
measurement is one of the main objectives of the future Electron
Ion Collider (EIC)~\cite{Boer:2010zf,Boer:2011fh,Zheng:2018ssm,xue:2021svd,Burkert:2022hjz,Abir:2023fpo} at Brookhaven , as well as of a similar facility being planned in 
China~\cite{Anderle:2021wcy}. 

There has been steady progress in the computation
of PDFs within lattice QCD in the last few years~\cite{Cichy:2018mum,Constantinou:2020pek}. 
Although many different approaches have been developed~\cite{Liu:1993cv,Aglietti:1998ur,Liu:1999ak,Detmold:2005gg,Braun:2007wv,Ma:2014jla,Ma:2014jga,Chambers:2017dov,Ma:2017pxb}, the two most widely used are those based on the computation of
quasi-PDFs~\cite{Ji:2013dva,Ji:2014gla,Ji:2020ect} and pseudo-PDFs~\cite{Orginos:2017kos,Radyushkin:2017cyf,Radyushkin:2019mye}. Extensive reviews of recent work on all of these approaches can be found in Ref.~\cite{Cichy:2018mum,Cichy:2021lih,Cichy:2021ewm,Constantinou:2022yye,Ji:2020ect}.

The former case is understood to lie within the framework of
the Large Momentum Effective theory (LaMET)~\cite{Ji:2013dva,Ji:2014gla,Ji:2020ect}.
LaMET allows one to compute not only the longitudinal 
PDFs but also GPDs~\cite{Ji:2015qla,Xiong:2015nua,Chen:2019lcm,Alexandrou:2020uyt,Lin:2020rxa,Alexandrou:2020zbe,Lin:2021brq} and
TMD~PDFs~\cite{Anselmino,Ji:2014hxa,Ji:2018hvs,Ji:2019sxk,Ebert:2022fmh,LatticePartonCollaborationLPC:2022myp}. 
The procedure is similar to that used for PDFs, but in this case, quasi-TMD PDFs are computed from purely spatial correlation functions, where the fermion fields are connected via a staple-shaped Wilson line rather than a straight line \cite{Ji:2014hxa,Ebert:2018gzl,Ji:2018hvs,Ji:2019ewn,Ebert:2019okf}.
These spatial correlators are evaluated with the hadron boosted to large momentum, and a matching procedure is then applied to extract the physical PDFs~\cite{Xiong:2013bka, Ishikawa:2016znu,Chen:2016fxx,Wang:2017qyg,Stewart:2017tvs,Alexandrou:2018yuy,Alexandrou:2018eet,Liu:2018hxv,LatticeParton:2018gjr,Alexandrou:2019lfo} and TMD~PDFs~\cite{Ji:2014hxa,Ebert:2018gzl,Ji:2018hvs,LatticePartonLPC:2023pdv,LatticeParton:2024mxp}. 
Unlike the case of PDFs, where the matching involves convolution integrals, the matching for TMD PDFs is simpler, being purely multiplicative. However, it involves two additional quantities that must be calculated on the lattice:
the rapidity-independent reduced soft function $S_r$~\cite{LatticeParton:2020uhz,Zhang:2020dbb,Li:2021wvl,LatticeParton:2023xdl} and the Collins–Soper kernel $K$~\cite{Ebert:2018gzl,Ebert:2019tvc,Shanahan:2019zcq,Shanahan:2020zxr,Shanahan:2021tst,Schlemmer:2021aij,LatticePartonLPC:2022eev,LatticeParton:2023xdl,Avkhadiev:2024mgd}. Both $S_r$ and $K$ can be extracted from purely spatial correlation functions within the LaMET framework. The perturbative matching kernel $H$, which connects quasi-TMD PDFs to light-front TMD PDFs, is calculated through perturbation theory~\cite{Ebert:2019okf,Ji:2018hvs}.

The reduced soft function is independent of the hadron state and thus, it can be extracted from the computation of a meson form factor involving a bilocal four-quark operator, evaluated using spatial correlation functions, together with the quasi-TMD wave functions (WFs)~\cite{Ji:2019sxk,LatticeParton:2023xdl}. This is done by combining two key factorizations within the LaMET framework: the matching between quasi-TMD and light-cone TMD WFs
and the factorization of the meson form factor in terms of quasi-TMD WFs.
The Collins–Soper kernel 
governs the rapidity evolution of TMD~PDFs. 
It can be computed on the lattice within the LaMET framework through ratios of either quasi-TMD PDFs or quasi-TMD WFs~\cite{Ji:2021znw} by using the corresponding factorization formula between quasi- and light-cone quantities. To date, only a few complete lattice QCD calculations of physical TMD PDFs have been presented~\cite{LatticePartonCollaborationLPC:2022myp,Bollweg:2025iol}.
An essential step in computing quasi-TMD quantities, such as quasi-TMD PDFs and WFs, is the renormalization of nonlocal quark bilinear operators containing a staple-shaped Wilson line that enter their definitions. The analysis of these operators is more challenging than in the straight Wilson line case. In addition to linear divergences from the finite staple length and logarithmic divergences at the endpoints, the staple geometry introduces additional logarithmic divergences at its cusps. Moreover, gluon exchange between the infinitely extended transverse segments leads to pinch-pole singularities. These operators also exhibit a large mixing pattern on the lattice~\cite{Alexandrou:2023ucc}, further complicating their renormalization analysis.
In this work, we closely follow the approach previously employed by us in Ref.~\cite{Alexandrou:2023ucc} to remove divergences and address operator mixing, which has also been adopted in Refs.~\cite{Ji:2021uvr, Zhang:2022xuw}.

The nonperturbative renormalization scheme employed is the Short Distance Ratio (SDR) scheme~\cite {LatticePartonCollaborationLPC:2022myp,Zhang:2022xuw}, in which the expectation values of appropriately constructed Wilson loops are used to cancel the linear, cusp, and pinch-pole singularities. Defined at short perturbative distances, this method is more suitable compared to the standard RI$'$/MOM scheme~\cite{Martinelli:1994ty}, which becomes unreliable at large separations due to substantial nonperturbative effects~\cite{Alexandrou:2023ucc} and residual divergences~\cite{Zhang:2022xuw}. This scheme is adopted after we show that the operator mixing predicted from symmetry arguments~\cite{Alexandrou:2023ucc} is  negligible.
    
 In this work, 	we employ two gauge ensembles, denoted cA211.53.24 and cA211.30.32, generated by the Extended Twisted Mass Collaboration (ETMC)~\cite{ETMC} using $\mathcal{N}_f = 2+1+1$ clover-improved twisted-mass fermions and the Iwasaki gauge action. Their parameters are given in Table~\ref{tab:params}. The two ensembles are used to calculate the quasi-TMD~WF 
 and the meson form factor. 
 The different volumes with the same lattice spacing of the two ensembles allow for the study of finite-volume effects. 

\begin{table}[h!]
\begin{tabular}{|c|c|c|c|c|c|c|}
\hline
	Ensemble name & $L^3\times T$ & $a$ (fm) & $a\mu_l$&$m_{\pi}$~(MeV) \\
	\hline
	cA211.53.24 & $24^3$ $\times$ 48 & 0.093 & 0.018 and 0.03 & 640 and 830\\
        \hline
	cA211.30.32 & $32^3$ $\times$ 64 & 0.093 & 0.0197 & 640\\
\hline 
	\end{tabular}
\caption{Parameters of the two ensembles analyzed. We give in the first column the ensemble name, in the second the lattice volume in lattice units, in the third the lattice spacing, in the fourth the light bare valence quark mass in lattice units, and in the fifth the valence pion mass $m_\pi$ in physical units.
\label{tab:params}}
\end{table}

The rest of the paper is organized as follows: In Section~\ref{sec:quasitmdwf}, 
we define the quasi-TMD~WF 
within the LaMET framework, along with the staple-shaped Wilson line and the vacuum expectation value of the associated Wilson loop that enter its definition.
Sections~\ref{sec:CSKernel} and~\ref{sec:soft} 
present the theoretical framework for the computation of the Collins-Soper kernel and of the reduced 
soft function in lattice QCD. In Section~\ref{sec:renormalization}, we give details on the
renormalization procedure, while in Section~\ref{sec:setup}, we present the analysis of the lattice QCD results. In Section~\ref{sec:results}, we present our results for the 
Collins-Soper kernel and the reduced soft function up to $\mathcal{O}\left(\alpha_s\right)$ corrections within LaMET, followed by 
the conclusions in Section~\ref{sec:conclusions}.
   
  \section{Quasi-transverse momentum dependent wave functions}	
  \label{sec:quasitmdwf}

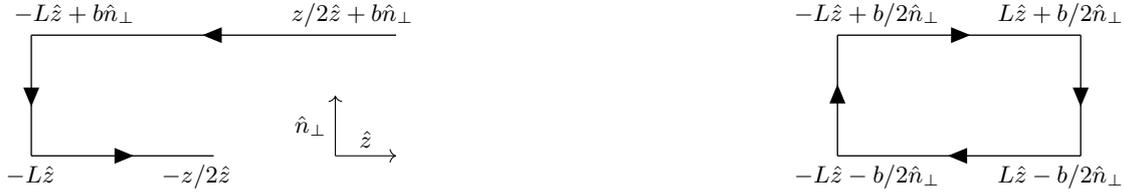
\begin{figure}[h!]
  \begin{minipage}[t]{0.42\linewidth}
  \begin{tikzpicture}[scale=0.8]
			\begin{feynhand}
				\vertex (j4)at (2,-1.0);
				\vertex (m)at (-1,-1);
				\vertex (n)at (-1,1);
				\vertex (j6)at (5,1);
				\vertex (x0)at (4,-1);
				\vertex (x1)at (5,-1);
				\vertex (x2)at (4,0);

				\propagator[antfer] (j4) to[edge label = $-L\hat{z}$ \qquad \qquad $-z/2\hat{z}$ \qquad] (m);
				\propagator[antfer] (m) to (n);
				\propagator[antfer] (n) to[edge label =$-L\hat{z}+b\hat{n}_{\perp}$ \qquad \qquad \qquad $z/2\hat{z}+b\hat{n}_{\perp}$] (j6);
				\draw [->] (x0) to[edge label=$\hat{z}$] (x1)[label=below];
				\draw [->] (x0) to[edge label=$\hat{n}_\perp$] (x2);
				
			\end{feynhand}
		\end{tikzpicture}
  \end{minipage}\hfill
  \begin{minipage}[t]{0.48\linewidth}
  \begin{tikzpicture}[scale=0.8]
		\begin{feynhand}
			\vertex (j4)at (2,-1.0);
			\vertex (m)at (-2,-1);
			\vertex (n)at (-2,1);
			\vertex (j6)at (2,1);

			\propagator[fer] (j4) to[edge label = $-L\hat{z}-b/2\hat{n}_{\perp}$ \qquad $L\hat{z}-b/2\hat{n}_{\perp}$] (m);
			\propagator[fer] (m) to (n);
			\propagator[fer] (n) to[edge label =$-L\hat{z}+b/2\hat{n}_{\perp}$ \qquad $L\hat{z}+b/2\hat{n}_{\perp}$] (j6);
			\propagator[fer] (j6) to (j4);
			
		\end{feynhand}
	\end{tikzpicture}
  \end{minipage}
  \caption{Graphical representation of the asymmetric staple (left) and Wilson loop (right).
	\label{fig:staple}} 
 \end{figure}

In the context of LaMET, the quasi-TMD~PDF $\tilde{f}_\Gamma$ is defined as the matrix element of a nonlocal fermion bilinear operator with a staple-shaped spatial Wilson line $\mathcal{W} (b,z,L)$: 
\begin{equation}
	\tilde{f}_\Gamma (x,b,\mu,\zeta_z) =  \lim_{L\to\infty} \int \frac{d z}{2 \pi} e^{-i z \zeta_z} \frac{P^z}{E_P} \langle N(P^z) \vert \bar{q}(y + b + z/2) \Gamma \mathcal{W}(b,z,L) q(y-z/2) \vert N(P^z) \rangle \mathcal{P},
\end{equation}
where $N(P^z)$ is a hadron state with momentum boost $P^z$ in the direction $\hat{z}$ and ground-state energy $E_P$. The Dirac matrix $\Gamma$ and the projector $\mathcal{P}$ determine the polarizations of the struck quark and the parent hadron, respectively, giving access to different TMD distributions~\cite{Ebert:2020gxr}. The Wilson line $\mathcal{W} (b,z,L)$ is formed by the product of spatial gauge links ordered in the shape of an asymmetric staple in the $(\hat{z}, \hat{n}_\perp)$ plane and is defined (for $y = 0$) as:
\begin{equation}
\mathcal{W}(b,z,L) = W_z^{\dagger}\left(-L\hat{z}+b\hat{n}_{\perp},(L+z/2)\right)W_{\perp}^{\dagger}\left(-L\hat{z},b\right)W_{z}\left(-L\hat{z},L-z/2\right), \label{Wilson line}
\end{equation}
where $W_z (y,l)$ ($W_\perp (y,l)$) is a straight Wilson line of length $l$, starting at the point $y$ and ending at the point $y+l \hat{z}$ ($y+l \, \hat{n}_\perp$). A graphical representation of $\mathcal{W}(z,b,L)$ is given in the left panel of Fig.~\ref{fig:staple}. 

In the large momentum limit, the light-cone TMD~PDF $f^{TMD}_\Gamma$ is
connected to the quasi-TMD-PDF 
    through the following relation~\cite{Ji:2020ect}:
    \begin{equation}
    f^{TMD}_\Gamma (x,b ,\mu,\zeta)=H\left(\frac{\zeta_z}{\mu^2}\right) e^{-\ln \left(\zeta_z/\zeta\right)K(b,\mu)}S^{1/2}_r (b,\mu)\tilde{f}_\Gamma (x,b,\mu,\zeta_z) + \mathcal{O}\left(\Lambda_{QCD}^2/\zeta_z,M^2/(P^z)^2,1/(b^2\zeta_z)\right),
    \label{eq:TMD_matching}
    \end{equation}
where 
     $S_r$ is the 
     rapidity-independent reduced soft function~\cite{LatticeParton:2020uhz,Li:2021wvl}, $H$ 
     is a matching kernel calculated in perturbation 
     theory~\cite{Ji:2020ect}, and $K$ is the Collins-Soper kernel~\cite{Ebert:2018gzl,Ebert:2019tvc,Shanahan:2019zcq}.
     Furthermore, $\zeta$ is the Collins-Soper scale that dictates the evolution of the rapidity, $\zeta_z=(2xP^z)^2$ is the corresponding scale
     in the case of the quasi-distributions, $b$ is the 
     distance in the transverse direction ($\hat{n}_\perp$) with respect to 
     the direction of the boost ($\hat{z}$), $\mu$ is the renormalization scale and $x$ is the momentum fraction carried by the parton. 
     Eq.~\eqref{eq:TMD_matching}  relates the physical  TMD~PDF to the quasi-TMD~PDF up to power corrections in $\Lambda_{QCD}^2/\zeta_z$, $M^2/(P^z)^2$, and $1/(b^2\zeta_z)$, with $M$ the mass of the target hadron.

As explained in sections~\ref{sec:CSKernel} and~\ref{sec:soft}, the Collins-Soper kernel and the reduced soft function can be expressed in terms of the quasi-TMD wave function (WF) $\tilde{\psi}_{\Gamma_{\phi}}$, defined in the coordinate space as:
\begin{eqnarray}
\tilde{\psi}_{\Gamma_{\phi}}\left(b,z,L,P^z\right) = \langle 0 | \bar{d}\left(t,y+b+z/2\right)\Gamma_{\phi}\mathcal{W}(b,z,L)u\left(t,y-z/2\right) | \pi\left(P^z\right)\rangle,\label{pion wavefunction}
\end{eqnarray}
where $|\pi\left( P^z \right)\rangle$ is the pion state with boost $P^z$, and $\Gamma_\phi$ is a Dirac gamma structure. To obtain the leading-twist contribution, $\Gamma_\phi$ is selected to be $\gamma_5 \gamma_3$ or $\gamma_5 \gamma_4$~\cite{Ji:2021znw},
where without loss of generality, the boost direction $\hat{z}$ is labeled as 3 and the temporal direction as 4. 

In line with standard practices, the quasi-TMD~WF is normalized using its local counterpart $\tilde{\psi}_{\Gamma_{\phi}}\left(0,0,0,P^z\right)$ evaluated at the same momentum $P^z$:
\begin{eqnarray}
\tilde{\psi}^{\rm norm.}_{\Gamma_{\phi}}\left(b,z,L,P^z\right) \equiv \frac{\tilde{\psi}_{\Gamma_{\phi}}\left(b,z,L,P^z\right)}{\tilde{\psi}_{\Gamma_{\phi}}\left(0,0,0,P^z\right)}. \label{norm wavefunction}
\end{eqnarray}
This normalization ensures that the perturbative matching coefficient to the light-cone TMD~WF (see Eq.~\eqref{eq:TMDWF_matching}) is one at leading order. An alternative normalization for the case $\Gamma_\phi = \gamma_5 \gamma_3$, which leads to the same perturbative matching coefficient, is:
\begin{eqnarray}
\tilde{\psi}^{\rm norm.}_{\gamma_5 \gamma_3} \left(b,z,L,P^z\right) \equiv \frac{\tilde{\psi}_{\gamma_5 \gamma_3}\left(b,z,L,P^z\right)}{\tilde{\psi}_{\gamma_5 \gamma_4}\left(0,0,0,P^z\right) \times (P^z / E_\pi)}, \label{norm wavefunction2}
\end{eqnarray}
where $E_\pi$ is the ground-state pion energy. Given that $\tilde{\psi}_{\gamma_5 \gamma_4} \left(0,0,0,P^z\right)$ generally provides a more stable signal, we employ Eq.~\eqref{norm wavefunction2} to determine the normalized quasi-TMD~WF for $\Gamma_\phi = \gamma_5 \gamma_3$.

As discussed in section~\ref{sec:renormalization}, the nonlocal operator $\bar{d}\left(t,y+b+z/2\right)\Gamma_{\phi}\mathcal{W}(b,z,L)u\left(t,y-z/2\right)$ entering $\tilde{\psi}_{\Gamma_{\phi}}$ suffers from linear and logarithmic divergences associated with the staple geometry. In order to cancel divergences that depend on the dimensions or the shape of the staple, namely linear, cusp and pinch-pole divergences (see section~\ref{sec:renormalization}), one often defines a subtracted quasi-TMD WF $\tilde{\psi}^{{\rm sub.},\,\pm}_{\Gamma_\phi}$ as:
\begin{eqnarray}
\tilde{\psi}^{{\rm sub.},\,\pm}_{\Gamma_{\phi}}\left(b,z,P^z\right) = \lim_{L\to\infty} \frac{\tilde{\psi}^{\rm norm.}_{\Gamma_{\phi}}\left(b,z,\pm L,P^z\right)}{\sqrt{Z_E\left(2 L, b\right)}},
\label{eq:psisub}
\end{eqnarray}
where $Z_E$ is the vacuum expectation value of a flat rectangular Euclidean Wilson loop~\cite{Collins:2011ca,Echevarria:2015byo,Ji:2018hvs,Shanahan:2019zcq} defined in the plane ($\hat{z}, \hat{n}_\perp$) with length $2L$ and 
width $b$, given by:
\begin{equation}
Z_E(2L, b) \!=\! \frac{1}{N_c}\!\biggl\langle\! W_{\perp}\!\left(-L\hat{z}-\frac{b}{2}\hat{n}_{\perp},b\right)\!W_z\!\left(-L\hat{z}+\frac{b}{2}\hat{n}_{\perp},2L\right) W_\perp^{\dagger}\!\left(L\hat{z}-\frac{b}{2}\hat{n}_{\perp},b\right)\!W_z^{\dagger}\!\left(-L\hat{z}-\frac{b}{2}\hat{n}_{\perp},2L\right)\biggl\rangle.
\label{ZE}
\end{equation}	
A graphical representation of $Z_E$ is given in the right panel of Fig.~\ref{fig:staple}. The cancellation of linear and pinch-pole divergences allows one to take the limit $L \rightarrow \infty$ 
 in order to match the light-cone definition of TMD~WF.

The momentum-space quasi-TMD~WF $\tilde{\Psi}^{\pm}_{\Gamma_{\phi}}$ is  defined as the Fourier transformation of $\tilde{\psi}^{{\rm sub.},\,\pm}_{\Gamma_{\phi}}$:
\begin{align}
&\tilde{\Psi}^{\pm}_{\Gamma_{\phi}}\left(x,b,\mu,\zeta_z\right) = \int_{-\infty}^{+\infty} \frac{P^z \ dz}{2\pi}e^{-\mathit{i}(x-\frac{1}{2})zP^z}Z(\mu)\,\tilde{\psi}^{{\rm sub.},\,\pm}_{\Gamma_{\phi}}\left(b,z,P^z\right) ,& 
\end{align}
where $x$ is the quark momentum fraction, and $Z (\mu)$ is a renormalization factor that eliminates the remaining divergences, namely the endpoint divergences (see section~\ref{sec:renormalization}). This factor is determined by employing short-range non-perturbative renormalization prescriptions, such as the RI-short~\cite{Ji:2021uvr} or Short Distance Ratio~\cite{LatticePartonCollaborationLPC:2022myp} schemes already explored in our previous  work for the quasi-TMD~PDFs~\cite{Alexandrou:2023ucc}. The two renormalization schemes are presented in section~\ref{sec:renormalization}.

As in the case of TMD PDFs (Eq.~\ref{eq:TMD_matching}), the quasi-TMD~WF $\tilde{\Psi}^\pm_{\Gamma_\phi}$ is connected to the light-cone TMD~WF $\Psi^\pm_{\Gamma_\phi}$ through the following matching relation~\cite{Ji:2021znw} within LaMET:
\begin{equation}
    \Psi^\pm_{\Gamma_{\phi}}(x,b,\mu,\zeta)=H^\pm_{\Psi}\left(\frac{\zeta_z}{\mu^2},\frac{\bar{\zeta}_z}{\mu^2}\right) e^{-\frac{1}{2}\ln \left(\frac{\mp\zeta_z+\mathit{i}\varepsilon}{\zeta}\right)K(b,\mu)}S^{1/2}_r (b,\mu)\tilde{\Psi}^\pm_{\Gamma_{\phi}}(x,b,\mu,\zeta_z) + \mathcal{O}\left(\Lambda_{QCD}^2/\zeta_z,M^2/(P^z)^2,1/(b^2\zeta_z)\right),
    \label{eq:TMDWF_matching}
\end{equation}
where $H^\pm_{\Psi}$ is a matching coefficient calculated in perturbation theory~\cite{Ji:2021znw}, and $\bar{\zeta}_z = (2(1-x)P^z)^2$. We will show in the next section how to extract the Collins-Soper kernel from ratios of quasi-TMD~WFs.

\section{Collins-Soper kernel}
\label{sec:CSKernel}

 To extract the Collins–Soper kernel $K(b,\mu)$, one can take ratios of quasi-TMD PDFs or WFs evaluated at different hadron momenta. In such ratios, the reduced soft function and the light-cone TMD quantities cancel, allowing $K(b,\mu)$ to be isolated and expressed in terms of the matching coefficients and the ratio itself. In this study, we use ratios of quasi-TMD WFs, which exhibit better signal quality on the lattice compared to quasi-TMD PDFs~\cite{Li:2021wvl, LatticePartonLPC:2022eev}. Namely, we take
 \begin{eqnarray}
\frac{\tilde{\Psi}^\pm_{\Gamma_{\phi}} (x,b,\mu,\zeta_{z1})}{\tilde{\Psi}^\pm_{\Gamma_{\phi}} (x,b,\mu,\zeta_{z2})} = \frac{H^\pm_{\Psi}\left(\frac{\zeta_{z2}}{\mu^2},\frac{\bar{\zeta}_{z2}}{\mu^2}\right) e^{-\frac{1}{2}\ln \left(\frac{\mp\zeta_{z2}+\mathit{i}\varepsilon}{\zeta}\right)K(b,\mu)}S^{1/2}_r (b,\mu)}{H^\pm_{\Psi}\left(\frac{\zeta_{z1}}{\mu^2},\frac{\bar{\zeta}_{z1}}{\mu^2}\right) e^{-\frac{1}{2}\ln \left(\frac{\mp\zeta_{z1}+\mathit{i}\varepsilon}{\zeta}\right)K(b,\mu)}S^{1/2}_r (b,\mu)} = e^{h^\pm_{\Psi} \left(\frac{\zeta_{z2}}{\mu^2},\frac{\bar{\zeta}_{z2}}{\mu^2}\right)-h^\pm_{\Psi}\left(\frac{\zeta_{z1}}{\mu^2},\frac{\bar{\zeta}_{z1}}{\mu^2}\right)+\frac{1}{2}\ln \left(\frac{\zeta_{z1}}{\zeta_{z2}}\right)K(b,\mu)},
\end{eqnarray}
where we used the next-to-leading order (NLO) expression for the matching coefficient $H^\pm_\Psi$  given by~\cite{Ji:2021znw}:
\begin{eqnarray}
H^\pm_{\Psi}\left(\frac{\zeta_{z}}{\mu^2},\frac{\bar{\zeta}_{z}}{\mu^2}\right) = e^{h^\pm_{\Psi} \left(\frac{\zeta_{z}}{\mu^2},\frac{\bar{\zeta}_{z}}{\mu^2}\right)},
\end{eqnarray}
with
\begin{eqnarray}
h^\pm_{\Psi}\left(\frac{\zeta_{z}}{\mu^2},\frac{\bar{\zeta}_{z}}{\mu^2}\right) = -\frac{\alpha_s}{4\pi}C_F\left(-4-\frac{5\pi^2}{6}+\left(l_\pm+\bar{l}_\pm\right)-\frac{1}{2}\left(l_\pm^2+\bar{l}_\pm^2\right)\right) + \mathcal{O}\left(\alpha_s^2\right),
\end{eqnarray}
and $l_\pm = \ln\left(\left(-\zeta_z \pm \mathit{i}\varepsilon\right)/\mu^2\right)$ and $\bar{l}_\pm = \ln\left(\left(-\bar{\zeta}_z \pm \mathit{i}\varepsilon\right)/\mu^2\right)$. Using the NLO result for $h^\pm_\Psi$, the Collins–Soper kernel can be extracted through:
\begin{eqnarray}
	K(b,\mu) = \frac{1}{\ln(P^z_1 / P^z_2)} \ln \left(\frac{\tilde{\Psi}^\pm_{\Gamma_{\phi}} (x,b,\mu,\zeta_{z1})}{\tilde{\Psi}^\pm_{\Gamma_{\phi}} (x,b,\mu,\zeta_{z2})}\right) - \frac{\alpha_s}{4\pi}C_F\left(4-\ln\left(\frac{\zeta_{z_1}\zeta_{z_2}\bar{\zeta}_{z_1}\bar{\zeta}_{z_2}}{\mu^8}\right) \mp 4i\pi\right) + \mathcal{O}\left(\alpha_s^2\right),
 \label{eq:kernel}
\end{eqnarray}
where $C_F=(N_c^2 - 1)/(2N_c)$, with $N_c=3$. The presence of an imaginary part in Eq.(\ref{eq:kernel}) is problematic, since the Collins–Soper kernel is a physical quantity and should be real. The origin of the nonzero imaginary part has recently been attributed to a linear infrared renormalon, which is canceled by power corrections of the form $1 / (b \ P^z)$ at each perturbative order in $\alpha_s$~\cite{Liu:2023onm}. As shown in Ref.~\cite{Avkhadiev:2024mgd}, this imaginary part is eliminated by subtracting the leading renormalon contributions~\cite{Liu:2023onm} at large $b$, and by incorporating $b$-dependent power corrections at small $b$~\cite{Avkhadiev:2023poz}. Based on these results, we can safely neglect the imaginary part in Eq. (\ref{eq:kernel}) and retain only the real part as our final result. 

\section{Reduced Soft function}
\label{sec:soft}

Within LaMET, the reduced soft function $S_{r}$ can be extracted as follows~\cite{Ji:2020ect,Ji:2019sxk}:
\begin{equation}
	S_{r}(b, \mu)= \left[\frac{Z_{\Gamma}\left(\mu\right)}{Z_{\Gamma_\phi}\left(\mu\right)}\right]^2 \, \frac{F_{\Gamma}^{\rm norm.}\left(b, P, P^{ \prime}, \mu\right)}{\mathcal{H}_{F_{\Gamma}}\left(b, P, P^{\prime}, \mu\right)},
    \label{eq:soft}
\end{equation}
where 
\begin{equation}
    F_{\Gamma}^{\rm norm.}\left(b, P, P^{\prime},\mu\right) \equiv \frac{F_{\Gamma}\left(b, P, P^{\prime}\right)}{N_{\Gamma}\,\tilde{\psi}_{\Gamma_{\phi}}\left(0,0,0,P^{\prime z}\right) \tilde{\psi}_{\Gamma_{\phi}}\left(0,0,0,P^z\right)},
    \label{Fnorm}
\end{equation}
and $F_{\Gamma}$ is a four-point meson form factor defined as:
\begin{equation}
	F_{\Gamma}\left(b, P, P^{\prime}\right)= \left\langle\pi\left(P^{\prime}\right)\left|\bar{u}(b\hat{n}_{\perp}) \Gamma u(b\hat{n}_{\perp}) \bar{d}(0) \Gamma d(0)\right| \pi(P)\right\rangle
    \label{F}
\end{equation}
with $P = \left(P^{0},0,0,P^{z}\right)$ and $P^{\prime} = \left(P^{0},0,0,-P^{z}\right)$. In accordance with the normalization of the quasi-TMD~WF (see Eq. \eqref{norm wavefunction}), the form factor $F_\Gamma$ is normalized as in Eq.~\eqref{Fnorm}. Note that, due to the symmetry properties explained in  Appendix~\ref{sec:symmetry}, we have $\tilde{\psi}_{\Gamma_{\phi}}\left(0,0,0,P^{\prime z}\right) = \pm \,\tilde{\psi}_{\Gamma_{\phi}}\left(0,0,0,P^z\right)$, where the upper (lower) sign corresponds to $\Gamma_\phi = \gamma_5 \gamma_4$ ($\Gamma_\phi = \gamma_5 \gamma_3$). $N_{\Gamma}$ depends on the Dirac matrix $\Gamma$ and ensures that the perturbative matching kernel $H^\pm_{F_\Gamma}$ given in Eq. \eqref{H_FG} is one at the leading order~\cite{Li:2021wvl}:
\begin{equation}
N_{\Gamma} = \left\{\begin{array}{rl}
1/(2N_c), & \Gamma = \openone,\gamma_{\perp}, \gamma_5\gamma_{\perp} \\
-1/(2N_c), & \Gamma = \gamma_5.
\end{array}\right .
\end{equation}
$Z_{\Gamma}\left(\mu\right)$ and $Z_{\Gamma_\phi}\left(\mu\right)$ are the renormalization factors of the local quark bilinear operators $\bar{q} (x) \Gamma q (x)$ and $\bar{q} (x) \Gamma_\phi q (x)$, respectively, calculated in Ref.~\cite{Li:2021wvl} using the standard RI$'$/MOM prescription. These factors renormalize the  form factor.  
$\mathcal{H}_{F_{\Gamma}}$ is defined as:  
\begin{eqnarray}
\mathcal{H}_{F_{\Gamma}}\left(b, P, P^{\prime}, \mu\right)=\int_{-\infty}^{\infty} d x \,\int_{-\infty}^{\infty} d x' \, H^\pm_{F_{\Gamma}}\left(x,x', P, P^{\prime}, \mu\right) \tilde{\Psi}^{\pm}_{\Gamma_{\phi}}{}^{\dagger}(x', b,\mu,\zeta_z ') \, \tilde{\Psi}^\pm_{\Gamma_{\phi}}(x, b,\mu,\zeta_z) ,
\label{calH_FG}
\end{eqnarray}
with $H^\pm_{F_\Gamma}$ being a matching kernel calculated in perturbation theory~\cite{Deng:2022gzi} and given by
\begin{eqnarray}
H^\pm_{F_{\Gamma}}\left(x,x', P, P^{\prime}, \mu\right) = 1 + \frac{\alpha_s}{4\pi}C_F\Biggl( h_0^{\Gamma}&+2\pi^2+\ln^2\left(-\frac{|x'|}{|x|}\mp\mathit{i}\varepsilon\right)+\ln^2\left(-\frac{|1-x'|}{|1-x|}\mp\mathit{i}\varepsilon\right)+& \nonumber\\&+h_1^{\Gamma}\ln\left(\frac{16|x||1-x||x'||1-x'|(P^z)^4}{\mu^4}\right)\Biggr) + 
\mathcal{O}\left(\alpha_s^2\right) ,&
\label{H_FG}
\end{eqnarray}
where  $h^{\openone}_0 = h^{\gamma_5}_0 = 4$, $h^{\gamma_{\perp}}_0 = h^{\gamma_5\gamma_{\perp}}_0 = -8 $, $h^{\openone}_1 = h^{\gamma_5}_1 = -2$, $h^{\gamma_{\perp}}_1 = h^{\gamma_5\gamma_{\perp}}_1 = 1$.
By substituting Eq.~\eqref{H_FG} into Eq.~\eqref{calH_FG},  
$\mathcal{H}_{F_{\Gamma}}$ takes the following form:
\begin{eqnarray}
\mathcal{H}_{F_{\Gamma}}\left(b, P, P^{\prime}, \mu\right)
&=&|J_1|^2\left\{1+\frac{\alpha_s}{4\pi}C_F\left[h_0^{\Gamma}+h_1^{\Gamma}\ln\left(\frac{16{P^z}^4}{\mu^4}\right)\right]\right\} \mp \alpha_s C_F \, {\rm Im}\Big[J_1^*\left(J_2+J_3\right)\Big] \nonumber \\
&& +2 \frac{\alpha_s}{4\pi}C_F\Big\{ {\rm Re} \Big[ J_1^*\left(h_1^{\Gamma}\left(J_2+J_3\right)+J_4+J_5\right)\Big]-|J_2|^2-|J_3|^2 \Big\} + \mathcal{O}\left(\alpha_s^2\right),
\label{eq:final_denominator}
\end{eqnarray}
where
\begin{eqnarray}
    &J_1=&\int_{-\infty}^{\infty} d x \, \tilde{\Psi}^\pm_{\Gamma_{\phi}}(x, b, \mu,\zeta_z) , \label{J1}\\
    &J_2=&\int_{-\infty}^{\infty} d x \, \tilde{\Psi}^\pm_{\Gamma_{\phi}}(x, b, \mu,\zeta_z)\ln\left(|x|\right) , \\
    &J_3=&\int_{-\infty}^{\infty} d x \, \tilde{\Psi}^\pm_{\Gamma_{\phi}}(x, b, \mu,\zeta_z)\ln\left(|1-x|\right) , \\
    &J_4=&\int_{-\infty}^{\infty} d x \, \tilde{\Psi}^\pm_{\Gamma_{\phi}}(x, b, \mu,\zeta_z)\ln\left(|x|\right)^2  ,\\
    &J_5=&\int_{-\infty}^{\infty} d x \, \tilde{\Psi}^\pm_{\Gamma_{\phi}}(x, b, \mu,\zeta_z)\ln\left(|1-x|\right)^2  \,. \label{J5}
\end{eqnarray}
The integrals in Eqs. (\ref{J1} -- \ref{J5}) can be computed using standard numerical methods.
We note that $\mathcal{H}_{F_{\Gamma}}\left(b, P^z, P^{z \prime}, \mu\right)$ is purely real, as expected. The sign in the term $\mp \alpha_s C_F \mathrm{Im}\left[J_1^*\left(J_2 + J_3\right)\right]$ in Eq.~\eqref{eq:final_denominator} is determined by the sign of the staple parameter $L$. However, this term is in fact independent of the sign of $L$, as the imaginary part of the quasi-TMD~WF changes sign when $L \to -L$, effectively canceling the overall sign in front of the term in Eq.~\eqref{eq:final_denominator}.

\section{NONPERTURBATIVE RENORMALIZATION}
\label{sec:renormalization}

Renormalizing nonlocal quark bilinear operators with staple-shaped Wilson lines presents significant challenges, primarily due to the various divergences introduced by the staple geometry (see, e.g., Ref.~\cite{Spanoudes:2024kpb} and references therein). These include power-law divergences that scale with the total length of the Wilson line, as well as logarithmic divergences arising at the singular points of the Wilson line, i.e., at the cusps (cusp divergences) and at the end points (endpoint divergences). In the limit of infinite lateral staple extent, additional complications arise from pinch-pole singularities~\cite{Ji:2018hvs} caused by gluon exchange between the two transverse segments. Moreover, operator mixing among a large set of Dirac structures~\cite{Alexandrou:2023ucc} is present on the lattice and must be carefully addressed.

In contrast to the case of local or ultralocal operators, the application of the standard RI$'$/MOM scheme in the renormalization of nonlocal operators is problematic at large distances due to the presence of residual linear divergences and significant nonperturbative effects~\cite{Zhang:2022xuw}. As examined in our previous work~\cite{Alexandrou:2023ucc}, there are two possible alternative renormalization schemes that can overcome the issues of RI$'$/MOM at a satisfactory level. The first scheme is a modified version of RI$'$/MOM, called RI$'$-short~\cite{Ji:2021uvr}, where amputated vertex functions of the staple operators are used divided by rectangular Wilson loops, which cancel divergences associated with the dimensions or the shape of the staple, i.e., power-law, cusp and pinch-pole divergences. The remaining endpoint divergences are eliminated through RI$'$/MOM-like conditions defined at short distances. The second scheme is a modified version of a ratio scheme, the SDR scheme~\cite{LatticePartonCollaborationLPC:2022myp,Zhang:2022xuw}, where appropriate ratios of hadron matrix elements of staple operators at different external momenta and at short distances are considered. The RI$'$-short scheme is the scheme of choice when there is non-negligible mixing among the staple operators, while SDR is only applicable when operator mixing is negligible. SDR is a gauge-invariant scheme and thus, there is no need to fix the gauge. On the contrary, RI$'$-short depends on the gauge. Given that we make use of Coulomb gauge-fixed momentum-wall source fields 
in the calculation of the quantities under study, the same gauge must be used in calculating the renormalization factors in the RI$'$-short scheme.   However, the conversion factors from RI$'$-short to the reference scheme of $\overline{\rm MS}$ are not available in the Coulomb gauge, and thus we restrict ourselves to the use of SDR scheme.  RI$'$-short is only considered for checking whether  mixing can be neglected so one can apply the SDR scheme.   

In what follows, we examine mixing within the RI$'$-short scheme. We define a subtracted amputated vertex function $\Lambda_{\Gamma}^{\rm sub.}$ as 
    \begin{equation}
    \Lambda_{\Gamma}^{\rm sub.}(b,z,p) = \lim_{L\to\infty} \frac{\Lambda_{\Gamma}(b,z,L,p)}{\sqrt{Z_E(2L, b)}},
    \end{equation}
    where 
    \begin{equation}
     {\Lambda_{\Gamma}}(b,z,L,p) = \langle S_u(p)^{-1} \, G_{\Gamma}(b,z,L,p) \, S_d(p)^{-1}\rangle, \label{vertex1}   
    \end{equation}
\begin{equation}
	{G_{\Gamma}}(b,z,L,p) = \sum_{{\bf x}_f,{\bf x}} e^{-\mathit{i} p \cdot (x_f-x_0)}
	\langle u(x_f) \, \bar{u}(x+b\hat{n}_{\perp}+z/2\hat{z})\Gamma\mathcal{W}(b,z,L) d(x-z/2\hat{z}) \, \bar{d}(x_0) \rangle,\label{vertex2}
	\end{equation}    
    and $S_u(p)$, $S_d(p)$ are the up and down quark propagators, respectively. The square root of the vacuum expectation value of the Wilson loop $Z_E$ cancels the divergences that depend on the staple parameters $b$ and $L$ that give rise to linear, cusp and pinch-pole singularities.\footnote{Note that the staple is constructed in such a way that there are no divergences depending on z.} Then, the limit of $L \rightarrow \infty$ can be taken. In the absence of such divergences, the vertex function is expected to exhibit a much milder dependence on the parameters $z$ and $b$. Therefore, we can fix the values of these parameters in the renormalization condition to have the small values~\cite{Ji:2021uvr}, $z_0$ and $b_0$, where perturbation theory works well. The RI$'$-short renormalization condition for the staple operators takes the following form:
    \begin{align}
    \left(Z_{\Gamma\Gamma'}^{{\rm RI}'{\rm -short}}(b_0,z_0,\mu_0)\right)^{-1} = \frac{{\rm Tr}\left[\Lambda_{\Gamma}^{\rm sub.}(b,z,\mu_0)\Gamma'^{\dagger}\right]}{12 e^{\mathit{i}\mathbf{\mu_0}\cdot\left(b\hat{n}_{\perp}+z\hat{z}\right)}Z_q^{{\rm RI}'{\rm /MOM}}(\mu_0)}\Big |_{z=z_0,b=b_0},
    \end{align}
    where
    \begin{equation}
    Z_q^{{\rm RI}'{\rm /MOM}}(\mu_0) = -\mathit{i}\frac{\rm Tr\left[\sum_{\mu}S(p)^{-1}\gamma_{\mu}\sin\left(p_{\mu}\right)\right]}{12\sum_{\mu}\sin^2\left(p_{\mu}\right)}\Big |_{p^{2}=\mu_0^2},
    \end{equation}
$\mu_0$ is the RI$'$-short renormalization scale, and $S(p)=(S_u(p)+S_d(p))/2$. The renormalization factor $Z_{\Gamma\Gamma'}^{{\rm RI}'{\rm -short}}(b_0,z_0,\mu_0)$ addresses the remaining endpoint divergences, as well as any mixing.

According to the symmetry arguments given in Ref.~\cite{Alexandrou:2023ucc}, staple-shaped quark bilinear operators of different Dirac structures mix in groups of 4 in the case of  actions that break chiral symmetry, as follows: $(\mathcal{O}_\Gamma, \mathcal{O}_{\Gamma\gamma_{\perp}},\mathcal{O}_{\Gamma\gamma_{z}}, \mathcal{O}_{\Gamma\gamma_{\perp}\gamma_{z}})$, where $\gamma_{\perp}$ ($\gamma_z$) is the gamma matrix corresponding to the perpendicular (longitudinal) direction $\hat{n}_{\perp}$ ($\hat{z}$) of the staple. In the case of a symmetric staple ($z=0$), the operators with the Dirac structure $\Gamma\gamma_{\perp}\gamma_{z}$ do not mix. Thus, the renormalization function $Z_{\Gamma\Gamma'}^{{\rm RI}'{\rm -short}}$ is at most a $4 \times 4$ matrix, and the renormalized quasi-TMD~WFs take the form:  
    \begin{align}
    \tilde{\psi}^{{\rm sub.},\,{\rm RI}'-{\rm short}}_{\Gamma_{\phi}}\left(b,z,P^z,\mu_0, b_0, z_0\right) = \sum_{\Gamma'_{\phi}}Z_{\Gamma_{\phi}\Gamma'_{\phi}}^{{\rm RI}'-{\rm short}}(b_0,z_0,\mu_0)\tilde{\psi}^{{\rm sub.}}_{\Gamma'_{\phi}}\left(b,z,P^z\right) \approx Z_{\Gamma_{\phi}\Gamma_{\phi}}^{{\rm RI}'-{\rm short}}(b_0,z_0,\mu_0)\tilde{\psi}^{{\rm sub.}}_{\Gamma_{\phi}}\left(b,z,P^z\right),
    \label{eq:ren}
    \end{align}
    where the approximation applies if operator mixing is negligible. 
   
\begin{figure}[h!]
    \centering
    \subfigure{\includegraphics[width=0.49\linewidth]{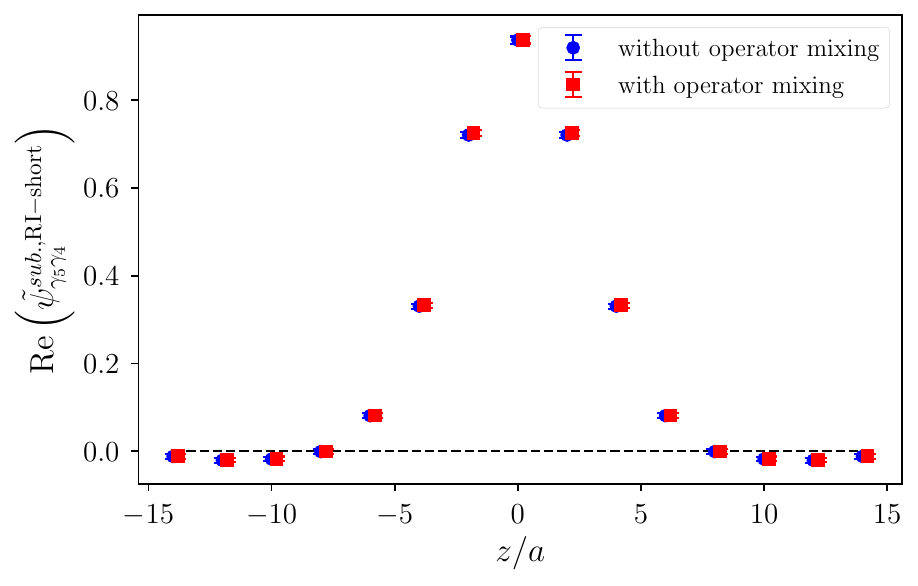}}%
    \subfigure{\includegraphics[width=0.49\linewidth]{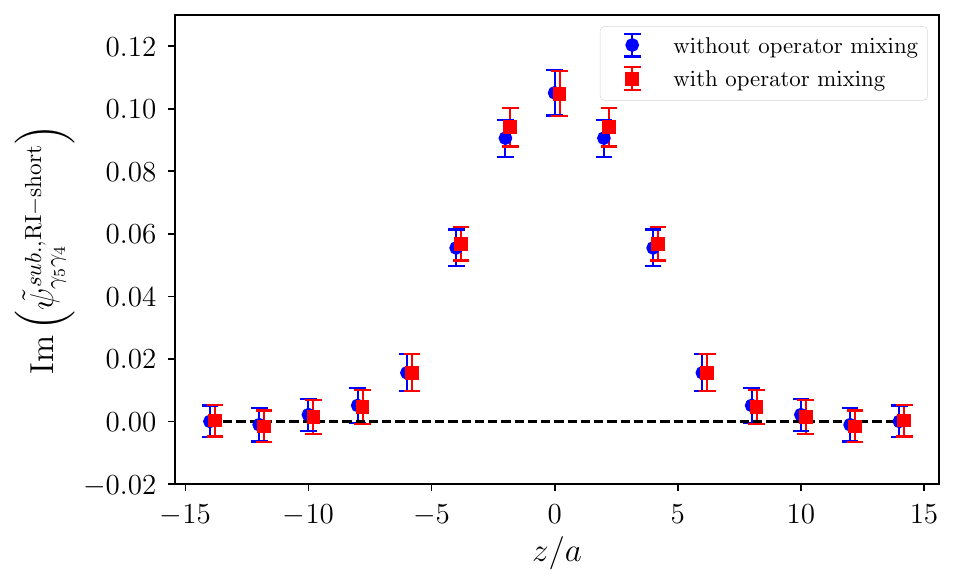}}
    \caption{Comparison of non-perturbatively renormalized results through the RI$'$-short scheme for the normalized and subtracted correlation functions $\tilde{\psi}_{\gamma_5\gamma_4}^{{\rm sub.}}$ with and without operator mixing at $b=2a$ and $L=8a$, using the cA211.53.24 ensemble and the valence pion mass $m_{\pi} = 830$~MeV. On the left, we show  the real  contribution and on the right the imaginary one.}
    \label{fig:mixing}
    \end{figure}

In Fig.~\ref{fig:mixing}, we compare the results of the renormalized subtracted quasi-TMD~WFs for the case where we use $\Gamma_\phi = \gamma_5 \gamma_4$ without taking into account mixing to those where the mixing is considered. The comparison is done for the staple parameter value $b=2a$ and  the cA211.53.24 ensemble with valence pion mass $m_{\pi} = 830$~MeV. As observed, the mixing contributions among staple-shaped Wilson-line operators are indeed negligible at the level of $\lesssim 3 \%$. This corroborates our findings for the proton quasi-beam function in Ref.~\cite{Alexandrou:2023ucc}. Moreover, our results align with lattice perturbation theory, where one-loop mixing effects are effectively canceled upon averaging over positive and negative values of the staple length $L$~\cite{Spanoudes:2024kpb}. Thus, we can safely ignore the mixing in what follows and consider multiplicative renormalization. This allows us to employ the SDR scheme as discussed below.

The SDR scheme is defined through finite ratios of subtracted quasi-TMD correlation functions at zero momentum and short distances $b_0$ and $z_0$. Given that the ultraviolet (UV) divergences of the subtracted and normalized quasi-TMD~WFs are independent of the Dirac structure $\Gamma_\phi$~\cite{Ebert:2019tvc,Deng:2022gzi,Spanoudes:2024kpb}, we consider the ratio ${\tilde{\psi}^{{\rm sub.}}}_{\Gamma_{\phi}}\left(b,z,P^z\right)/{\tilde{\psi}^{{\rm sub.}}}_{\gamma_5 \gamma_4} \left(b_0,z_0,0\right)$ and determine the $\overline{\rm MS}$-renormalized quasi-TMD~WF through the following relation:
\begin{equation}
    \tilde{\psi}^{{{\rm sub.}},\,\overline{\rm MS}}_{\Gamma_{\phi}}\left(b,z,P^z,\mu\right) = \frac{{\tilde{\psi}^{{\rm sub.}}}_{\Gamma_{\phi}}\left(b,z,P^z\right)}{{\tilde{\psi}^{{\rm sub.}}}_{\gamma_5 \gamma_4} \left(b_0,z_0,0\right)}     [\tilde{\psi}^{{{\rm sub.}},\,\overline{\rm MS}}_{\gamma_5 \gamma_4} \left(b_0,z_0,0,\mu\right)]_{\rm pert},
    \label{eq:MS}
\end{equation}
where~\cite{Zhang:2022xuw} 
\begin{equation}
    [\psi^{{{\rm sub.}},\,\overline{\rm MS}}_{\gamma_5 \gamma_4} \left(b_0,z_0,0,\mu\right)]_{\rm pert.}=1+\frac{\alpha_sC_F}{2\pi}\left(\frac{1}{2}+\frac{3}{2}\ln\left(\frac{b_0^2+z_0^2}{4e^{-2\gamma_E}}\mu^2\right)-2\frac{z_0}{b_0}\textrm{arctan}\frac{z_0}{b_0}\right) + \mathcal{O}(\alpha_s^2).
    \label{conversion}
\end{equation}
The ratio ${\tilde{\psi}^{{\rm sub.}}}_{\Gamma_{\phi}}\left(b,z,P^z\right)/{\tilde{\psi}^{{\rm sub.}}}_{\gamma_5 \gamma_4} \left(b_0,z_0,0\right)$ is UV-finite and, thus, directly comparable to its $\overline{\rm MS}$ counterpart. Moreover, ratios of this kind help mitigate certain systematic uncertainties. In principle, when different Dirac structures are used in the numerator and denominator ($\Gamma_\phi \neq \gamma_5 \gamma_4$), the bare lattice and $\overline{\rm MS}$-renormalized ratios can differ by UV-finite terms arising from the corresponding ratio of multiplicative renormalization factors in $\overline{\rm MS}$. However, as concluded by lattice perturbation theory, these finite terms are independent of the Dirac structure, at least to one-loop order~\cite{Spanoudes:2024kpb}, and thus cancel in the ratio. 
Furthermore, we cannot employ $\Gamma_\phi = \gamma_5 \gamma_3$ in the denominator of the ratio because the normalization constant $\tilde{\psi}_{\gamma_5 \gamma_3} (0,0,0,0)$ (see Eq. \eqref{norm wavefunction}) vanishes at zero momentum leading to a singular  ratio. %

\begin{figure}[h!]
    \centering
    \subfigure{\includegraphics[width=0.49\linewidth]{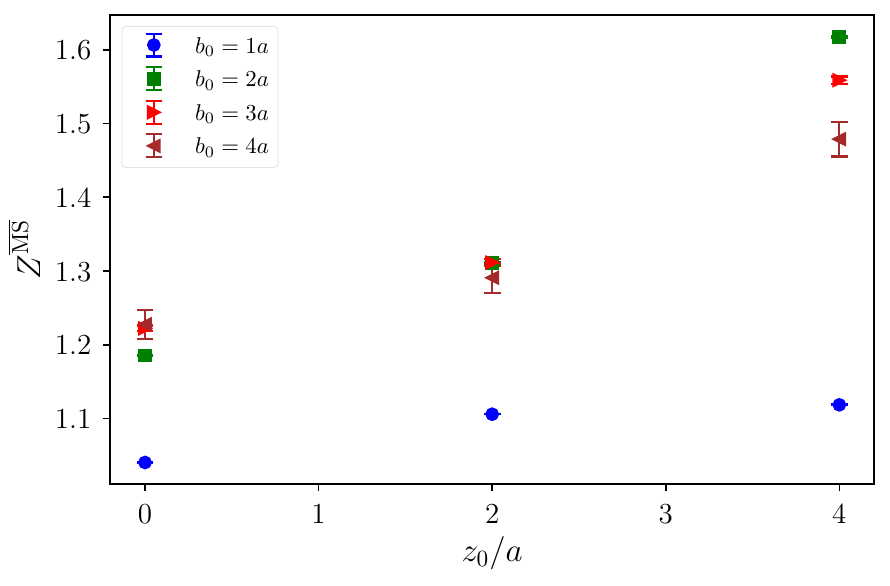}}%
    
    \caption{Perturbative renormalization factor calculated through SDR method.  The results are obtained with $L=8a$ using the cA211.53.24 ensemble and the valence pion mass $m_{\pi} = 830$~MeV.
    }
    \label{fig:B0}
\end{figure}

 According to Eq.\eqref{eq:MS}, the renormalization factor $Z^{\overline{\rm MS}} (\mu)$ for ${\tilde{\psi}^{{\rm sub.}}}_{\Gamma_{\phi}}\left(b,z,P^z\right)$ is determined by:
\begin{equation}
    Z^{\overline{\rm MS}} (\mu) = \frac{[\psi^{{{\rm sub.}},\,\overline{\rm MS}}_{\gamma_5 \gamma_4} \left(b_0,z_0,0,\mu\right)]_{\rm pert.}}{{{\tilde{\psi}^{{\rm sub.}}}_{\gamma_5 \gamma_4} \left(b_0,z_0,0\right)}}.
\end{equation}
Although the values of $z_0$ and $b_0$ can, in principle, be chosen arbitrarily, they should be small enough to ensure the validity of perturbation theory. In this work, we explore different choices of $b_0$ and $z_0$, selecting the values for which $Z^{\overline{\rm MS}} (\mu)$  exhibits good convergence. We find that this occurs at $b_0 = 3a$ and $z_0 = 0$, as illustrated in Fig. \ref{fig:B0}. The results shown are  obtained with $L=8a$ using the cA211.53.24 ensemble and the valence pion mass $m_{\pi} = 830$~MeV.

\section{ANALYSIS OF LATTICE QCD CORRELATORS}
\label{sec:setup}

The quasi-TMD~WF is computed using momentum boosts in all spatial directions and staple-shaped Wilson-line quark bilinear operators in the corresponding transverse directions, employing both positive and negative values of the staple parameters $b$, $z$, and $L$. For each configuration, we use $N_{{\rm sources}}$ momentum wall-source fields. The source-sink time separation $t_s$ is kept fixed, while the source time $t_0$ and sink time $t_f$ are shifted accordingly. As a result, for each configuration we obtain a total of $ N_{{\rm sources}} \times (6 \ \text{directions of momentum}) \times (16 \ \text{independent staple-shaped Wilson lines})$ measurements. The 16 Wilson lines arise from two transverse directions and two choices (positive and negative) for each of the three staple parameters. We exploit the symmetry properties of the quasi-TMD~WF given in Appendix~\ref{sec:symmetry} to combine all these measurements, reducing statistical noise. The values of $t_s$ and the momentum boosts used for each ensemble are given in Table~\ref{tab:momenta}. The total number of measurements $N_{{\rm stat}} = N_{{\rm confs}} \times N_{{\rm sources}} \times 96$ for each ensemble is given in Table~\ref{tab:statistics}.
 
\begin{table}[h!]
 \begin{tabular}{|c|c|c|c|}
\hline
	Ensemble name & $aP^z$ & $|P^z|$~(GeV)& $t_{s}/a$ \\
	\hline
    cA211.53.24 & \makecell[c]{$ \{ \pm 4,0,0 \}2\pi/24$ \  $ \{ \pm 6,0,0 \}2\pi/24 $ \\ $ \{ 0,\pm 4,0 \}2\pi/24$ \  $ \{ 0,\pm 6,0 \}2\pi/24 $ \\ $ \{ 0,0,\pm 4 \}2\pi/24$ \  $ \{ 0,0,\pm 6 \}2\pi/24 $}& 2.22 and 3.33 & 6, 8, 10 \\
    \hline
    cA211.30.32 & \makecell[c]{$ \{ \pm 4,0,0 \}2\pi/32$ \  $ \{ \pm 6,0,0 \}2\pi/32 $ \\ $ \{ 0,\pm 4,0 \}2\pi/32$ \  $ \{ 0,\pm 6,0 \}2\pi/32 $ \\ $ \{ 0,0,\pm 4 \}2\pi/32$ \  $ \{ 0,0,\pm 6 \}2\pi/32 $}&  1.63  and 2.45 & 6, 8, 10  \\
\hline 
	\end{tabular}
  \caption {We give in the first column the ensemble name. For the momentum boosts $P^z$ we give their values in lattice units (second column) and in physical units (third column). In the fourth column, we give the source-sink separations in lattice units. \label{tab:momenta}}
\end{table}

\begin{table}[h!]
 cA211.53.24 \\
\vspace{0.05cm} \begin{tabular}{|c|c|c|}
\hline
	$N_{\rm confs}$ & $N_{\rm sources}$ & $N_{\rm stat}$ \\
	\hline
     \makecell[c]{$ 500 \ \text{for} \ m_{\pi} = 830 \ \rm{MeV}, P^z = 2.22 \ \rm{GeV}$ \\ $1500 \ \text{for} \ m_{\pi} = 830 \ \rm{MeV}, P^z = 3.33 \ \rm{GeV}$ \\ $1200 \ \text{for} \ m_{\pi} = 640 \ \rm{MeV} $} & \makecell[c]{$ 24 \ \text{for} \ m_{\pi} = 830 \ \text{MeV}$ \\ $20 \ \text{for} \ m_{\pi} = 640 \ \text{MeV} $} & \makecell[c]{$  1152000 \ \text{for} \ m_{\pi} = 830 \ \rm{MeV}, P^z = 2.22 \ \rm{GeV}$ \\  $3456000 \ \text{for} \ m_{\pi} = 830 \ \rm{MeV}, P^z = 3.33 \ \rm{GeV}$ \\ $2304000 \ \text{for} \ m_{\pi} = 640 \ \rm{MeV} $} \\
    \hline
	\end{tabular} \\
\mbox{}\\
cA211.30.32\\
\vspace{0.05cm}\begin{tabular}{|c|c|c|}
\hline
$N_{\rm confs}$ & $N_{\rm sources}$ & $N_{\rm stat}$ \\
\hline
800 & 30 & 2304000 \\
\hline 
\end{tabular}
  \caption {The  parameters of the analysis are given  for the cA211.53.24 ensemble (top) and  for the cA211.30.32 ensemble (bottom). We give in the first column the number of configurations, in the second column  the number of momentum wall-sources per configuration and in the third column the total statistics.
  \label{tab:statistics}}
\end{table}

In order to extract the quasi-TMD~WF, we compute two-point correlation functions consisting of a pion interpolating operator and a staple-shaped Wilson-line quark bilinear operator. To obtain the subtracted version, we also calculate the vacuum expectation value of the Wilson loop defined in Eq. \eqref{ZE}. Furthermore, to extract the meson form factor, we evaluate four-point functions involving two local quark bilinear insertions. The staple-shaped Wilson lines used in both quark bilinear operator and the Wilson loop are stout-smeared~\cite{Morningstar:2003gk} using $n_{\rm stout}=5$ smearing steps and stout parameter $\rho_{\rm stout}=0.1315$.

  The pion interpolating field is given by 
  \begin{align}
  J_{\pi}(t;P^z) = \sum_{\mathbf{x},\mathbf{y}}e^{-\mathit{i} {\bf P}.{\bf x}/2} e^{-\mathit{i} {\bf P}.{\bf y}/2}\bar{d}\left(\mathbf{x},t\right) \gamma_5 u\left(\mathbf{y},t\right) , 
  \end{align} 
 where $\mathbf{x}$ and $\mathbf{y}$ represent the spatial coordinates of two independent momentum wall-sources. We note that the pion momentum is  symmetrically distributed between the source and the sink. This leads to a more symmetric definition of the two-point function, whose symmetry properties are detailed in Appendix~\ref{sec:symmetry}. These symmetries are exploited to enhance the statistical precision of our results by averaging over equivalent two-point functions.
The two-point and four-point correlation functions are given by
	\begin{eqnarray}
        C_{wf}\left(t,P^z,\Gamma_{\phi};b,z,L\right) &=& a^3\sum_{\mathbf{x}} e^{-\mathit{i}\bf{P}. \bf{x} }\langle \mathcal{O}_{u\bar{d}}^{\Gamma_{\phi}}\left(x;b,z,L\right)J_{\pi}^\dagger(t_0;P^z)\rangle,  \label{eq:twop}\\
	C_{4pt}\left(t,t_{s},P^z, \Gamma;b\right) &=& a^3\sum_{\mathbf{x}} e^{-2\mathit{i}\bf{P}.\bf{x}} 
	\langle  J_{\pi}(t_f;-P^z)\left(\bar{u}\Gamma u\right)\left(x+b\hat{n}_{\perp}\right)\left(\bar{d}\Gamma d\right)\left(x\right)J_{\pi}^\dagger(t_0;P^z) \rangle, \label{eq:fourp}
	\end{eqnarray}	
 where $t_0$ denotes the wall-source and $t_f$ the wall-sink, $t_{s}\equiv t_f-t_0$ is the time separation between the source and the sink, $t$ is the insertion time of the non-local quark bilinear operator: 
\begin{equation}
\mathcal{O}_{u\bar{d}}^{\Gamma_{\phi}}\left(x;b,z,L\right) = \bar{d}\left(t,x+b+z/2\right)\Gamma_{\phi}\mathcal{W}(b,z,L)u\left(t,x-z/2\right).
\end{equation}
The spectral decompositions of the correlation
functions are given by:
    \begin{eqnarray}
C_{wf}\left(t,P^z,\Gamma_{\phi};b,z,L\right) &=& \sum_{n=0}^{\infty} \frac{Z^{(n)}\left(P^z\right)\tilde{\psi}^{(n)}_{\Gamma_{\phi}}\left(b,z,L;P^z\right)}{2E^{{(n)}}\left(P^z\right)} e^{-E^{{(n)}}(P^z)t}, \label{eq:Cwf}\\
    C_{4pt}\left(t,t_{s},P^z,\Gamma;b\right)  &=& 
    \sum_{n,m=0}^{\infty} \frac{Z^{(n)}\left(-P^z\right) Z^{(m)}\left(P^z\right) F_\Gamma^{(n,m)}\left(b;P^z\right)}{4 E^{(n)}\left(P^z\right) E^{(m)}\left(P^z\right)} e^{-E^{(n)}(P^z)t_s} e^{[E^{(n)}(P^z) - E^{(m)}(P^z)] \, t},
    \label{eq:C4pt}
    \end{eqnarray}
where $\psi_{\Gamma_{\phi}}^{(n)}\left(b,z,L;P^z\right) = \langle 0|\mathcal{O}_{u\bar{d}}^{\Gamma_{\phi}}\left(0;b,z,L\right) |\pi^{(n)}(P^z)\rangle$ is the staple-shaped operator matrix element associated with the n$^{\rm th}$ excited state of the pion, $F_{\Gamma}^{(n,m)}\left(b;P^z\right) = \langle \pi^{(n)}(-P^z)|\left(\bar{u}\Gamma u\right)\left(0\right)\left(\bar{d}\Gamma d\right)\left(0\right)|\pi^{(m)}(P^z)\rangle$, is the form factor matrix element associated with the n$^{\rm th}$ and m$^{\rm th}$ excited states of the pion, $E^{(n)}(P^z)$ is the energy of the n$^{\rm th}$ excited state and $Z^{(n)}(P^z) = \langle J_{\pi} (0;P^z)|\pi^{(n)}(P^z)\rangle$ is the 
overlap of the correlation function with the n$^{\rm th}$ excited state.
To extract the normalized quasi-TMD~WF and the normalized form factor, we consider the following ratios: 
  \begin{eqnarray} 
    C_{wf}^0\left(t,P^z,\Gamma_{\phi};b,z,L\right) &\equiv& \frac{C_{wf}\left(t,P^z,\Gamma_{\phi};b,z,L\right)}{C_{wf}\left(t,P^z,\gamma_5 \gamma_4;0,0,0\right) R_{\Gamma_\phi}} \label{eq:Cwf0}\\
    C_{4pt}^0\left(t,t_s,P^z,\Gamma;b\right) &\equiv& \frac{C_{4pt}\left(t,t_{s},P^z,\Gamma;b\right)}{C_{wf}\left(t_s/2,-P^z,\Gamma_{\phi};0,0,0\right) C_{wf}\left(t_s/2, P^z,\Gamma_{\phi};0,0,0\right)},
    \label{eq:C4pt0}
    \end{eqnarray}
where $R_{\gamma_5 \gamma_4} = 1$ and $R_{\gamma_5 \gamma_3} = P^z/E^0 (P^z)$.    
The ground-state contributions of $C^0_{wf}$ and $C^0_{4pt}$, as obtained from the spectral decomposition of Eqs.\eqref{eq:Cwf} and \eqref{eq:C4pt}, give the bare normalized quasi-TMD~WF:
\begin{equation}
\tilde{\psi}^{(0)}_{\Gamma_{\phi}}\left(b,z,L;P^z\right) / \left[\tilde{\psi}^{(0)}_{\gamma_5 \gamma_4}\left(0,0,0;P^z\right) R_{\Gamma_\phi}\right] = \tilde{\psi}^{\rm norm.}_{\Gamma_{\phi}}\left(b,z,L,P^z\right),
\end{equation}
and the bare normalized form factor (up to a factor of $N_\Gamma$):  
\begin{equation}
F_\Gamma^{(0,0)} (b;P^z) / \left[\tilde{\psi}^{(0)}_{\Gamma_{\phi}} (0,0,0;-P^z) \,\tilde{\psi}^{(0)}_{\Gamma_{\phi}} (0,0,0;P^z) \right] = N_\Gamma \, F_\Gamma^{\rm norm.} (b,P^z).
\end{equation}
In our analysis, we perform both one-state and two-state fits and check the convergence of the extracted quantities. The fitting procedure is described in detail in Appendix~\ref{sec:fitting}. 

We employ $\Gamma_{\phi}=\gamma_5\gamma_4$ and $\Gamma_{\phi}=\gamma_5\gamma_3$ in $C_{wf}$, which give the leading-twist contributions to the quasi-TMD~WF~\cite{Ji:2021znw}. Since operator mixing is negligible, as shown in Section~\ref{sec:renormalization}, we do not employ any other Dirac structures. Furthermore, we employ $\Gamma = \gamma_5 \gamma_\perp, \gamma_\perp, \gamma_5 \gamma_j, \gamma_j$ in $C_{4pt}$ (where $j$ is orthogonal to both boost and transverse directions of the staple) and take appropriate combinations to isolate the leading-twist contributions to the form factor~\cite{Li:2021wvl} as explained in  Section~\ref{sec:results_soft}. 

After extracting the bare matrix elements, we apply the renormalization procedure as outlined in the previous sections. For the quasi-TMD~WF, in addition to dividing by the expectation value of a Wilson loop, the multiplicative renormalization factor in the SDR prescription is obtained using the ratio $C^0_{wf}$ at zero momentum (cf. Eqs.(\ref{eq:MS},\ref{eq:Cwf0})): 
\begin{align}
   \frac{1}{C_{wf}^{0}\left(t,P^z=0,\Gamma_{\phi} = \gamma_5 \gamma_4;b_0,z_0,L\right)} = \frac{C_{wf}\left(t,0,\gamma_5 \gamma_4;0,0,0\right)}{C_{wf}\left(t,0,\gamma_5 \gamma_4;b_0,z_0,L\right)} \ ,
\end{align}
which isolates the normalized ground-state element
\begin{equation}
\frac{1}{\tilde{\psi}^{\rm norm}_{\gamma_5 \gamma_4} \left(b_0,z_0,L,P^z=0\right)} = \frac{\tilde{\psi}^{(0)}_{\gamma_5 \gamma_4}\left(0,0,0;0\right) }{ \tilde{\psi}^{(0)}_{\gamma_5 \gamma_4}\left(b_0,z_0,L;0\right)}. 
    \end{equation}
In the following, we demonstrate that the two leading-twist Dirac structures, $\Gamma_\phi = \gamma_5\gamma_3$ and $\Gamma_\phi = \gamma_5\gamma_4$, yield compatible results for the quasi-TMD~WFs at large momenta, confirming the findings of Ref.~\cite{LatticePartonLPC:2022eev}. Following the analysis  of our previous study, we proceed to extract the Collins–Soper kernel and reduced soft function using $\Gamma_{\phi}=\gamma_5\gamma_4$.

\section{Results}
\label{sec:results}

\subsection{Quasi-Transverse Momentum Dependent Wave Functions}
\label{sec:results_tmdwf}

In this section, we present our results for the quasi-TMD~WF computed using the setup described in Section~\ref{sec:setup}. The results correspond to the Dirac structures $\Gamma_{\phi} = \gamma_5 \gamma_3$ and $\Gamma_{\phi} = \gamma_5 \gamma_4$, after averaging over all staple orientations and boost directions, as dictated by symmetries. 

\begin{figure}[h!]
    \centering
    \subfigure{\includegraphics[width=0.49\linewidth]{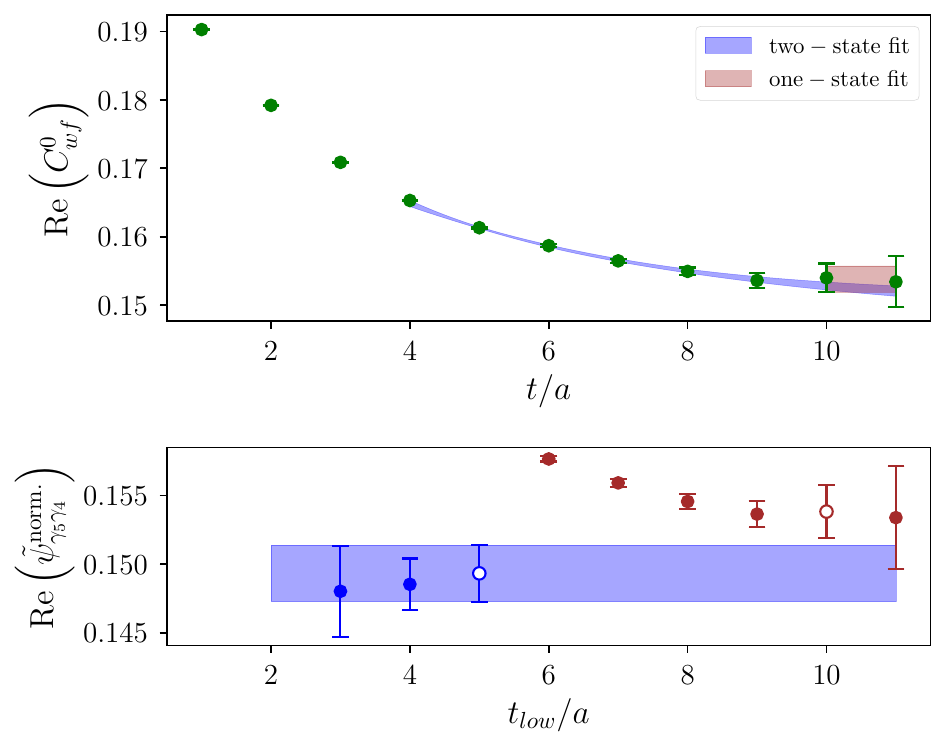}}%
    \subfigure{\includegraphics[width=0.49\linewidth]{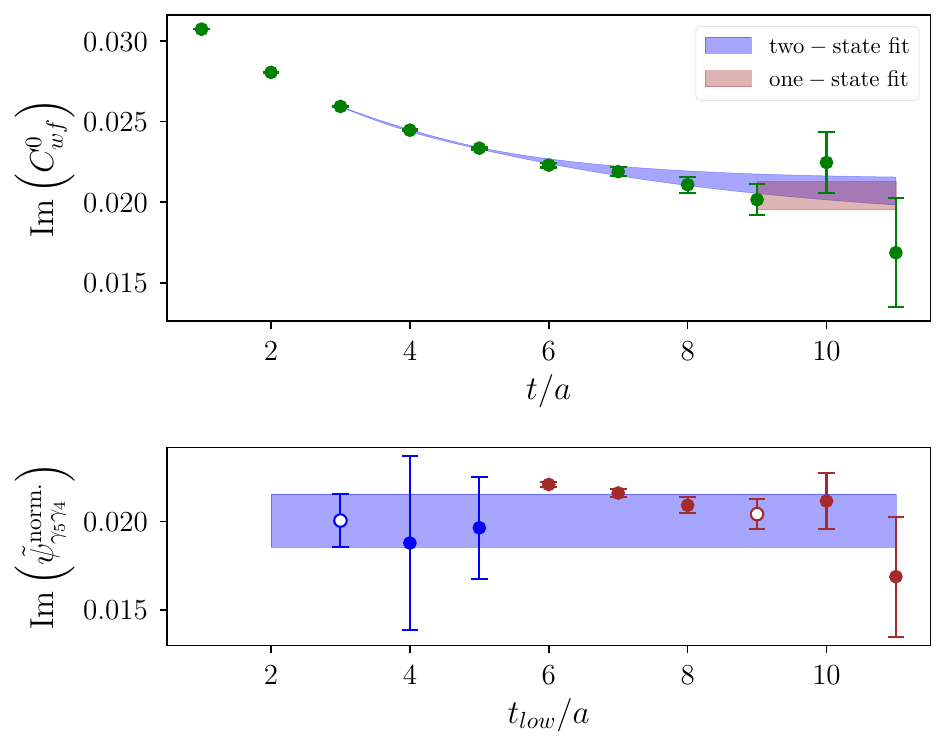}}\\
    \subfigure{\includegraphics[width=0.49\linewidth]{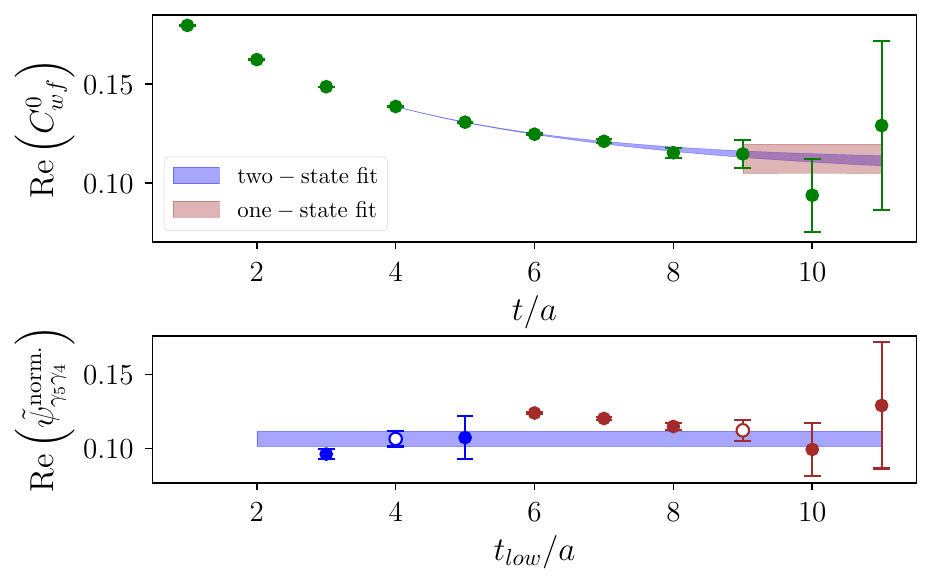}}%
    \subfigure{\includegraphics[width=0.49\linewidth]{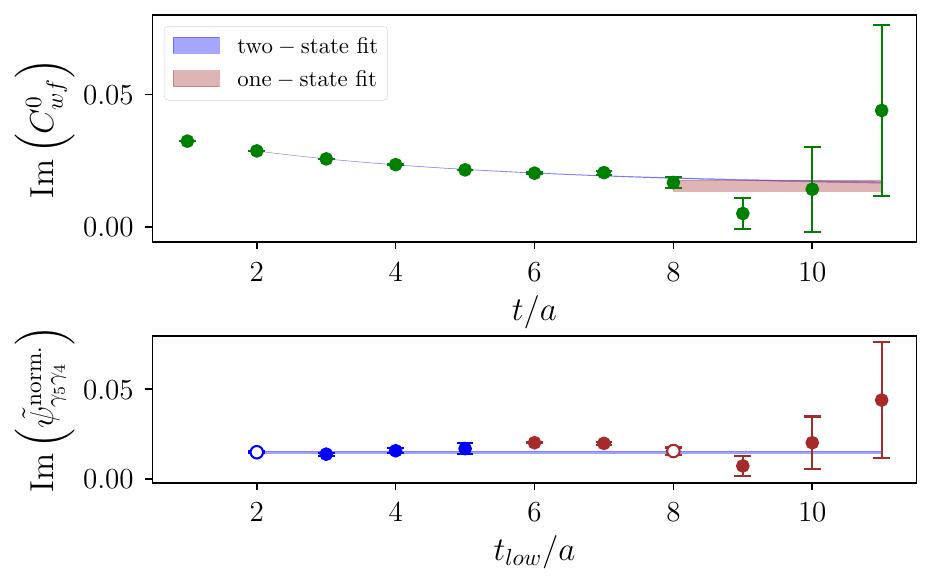}}
    \caption{Comparison of the one-state and two-state fits for the normalized correlation functions $C_{wf}^0$ employing the Dirac structure $\Gamma_{\phi} = \gamma_5\gamma_4$ for the cA211.53.24 ensemble and the valence pion mass $m_{\pi} = 830$~MeV. The plots correspond to the staple parameters $b=2a$ and $z=2a$ at boosts $P^z = 2.22$~GeV (top two rows) and $P^z = 3.33$~GeV (last two rows). The first and third rows show the real (left) and imaginary (right) parts of $C_{wf}^0$ for each momentum, along with the corresponding one- and two-state fitting bands. The second and fourth rows show results for the ground-state $\psi_{\gamma_5\gamma_4}^{\rm norm}$, for each momentum, as a function of $t_{\rm low}$, extracted from both one-state (in brown) and two-state (in blue) fits. The final selected values for the two fits are denoted with open symbols. The blue band in the second and fourth rows correspond to the selected two-state result used in the remainder of the analysis. }
    \label{fig:fitting}
\end{figure}

In Fig.~\ref{fig:fitting}, we show the extraction of the ground-state contribution to the ratio $C^0_{wf}$ using both one-state and two-state fits. The results are obtained for the cA211.53.24 ensemble and the valence pion mass $m_{\pi} = 830$~MeV and they correspond to boosts $P^z = 2.22$~GeV and $P^z = 3.33$~GeV, with staple parameters $b=2a$, $z=2a$, and $L=8a$. Larger values of $z$ lead to a rapid suppression of the  signal, particularly at the higher boost. We present both the real and imaginary parts of the correlation function.
To ensure stability of the ground state extraction, we vary the lower bound of the fit range, $t_{\rm low}$, and monitor convergence. For the one-state fit, convergence is observed for $t/a\gtrsim 8$. As our final result for the normalized ground-state matrix element, $\psi^{\rm norm}_{\Gamma_{\phi}}\left(b,z,L;P^z\right)$, we take the value from the two-state fit that agrees with the one-state fit within one standard deviation. Similar results are also obtained for the other leading-twist contribution, $\Gamma_{\phi} = \gamma_5\gamma_3$. A similar picture is observed across the other pion masses studied in this work, as well as for the cA211.30.32 ensemble.
\begin{figure}[hp!]
    \centering
    \subfigure{\includegraphics[width=0.49\linewidth]{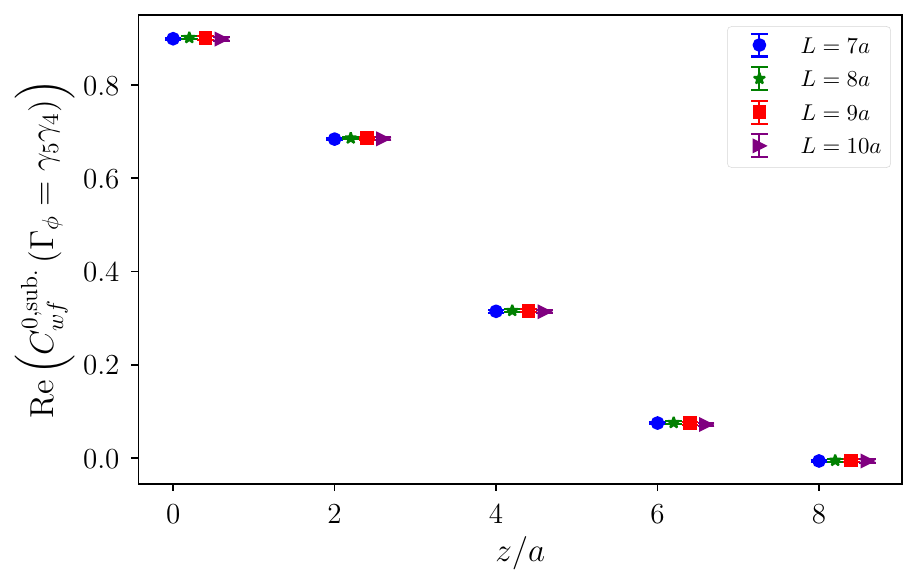}}%
    \subfigure{\includegraphics[width=0.49\linewidth]{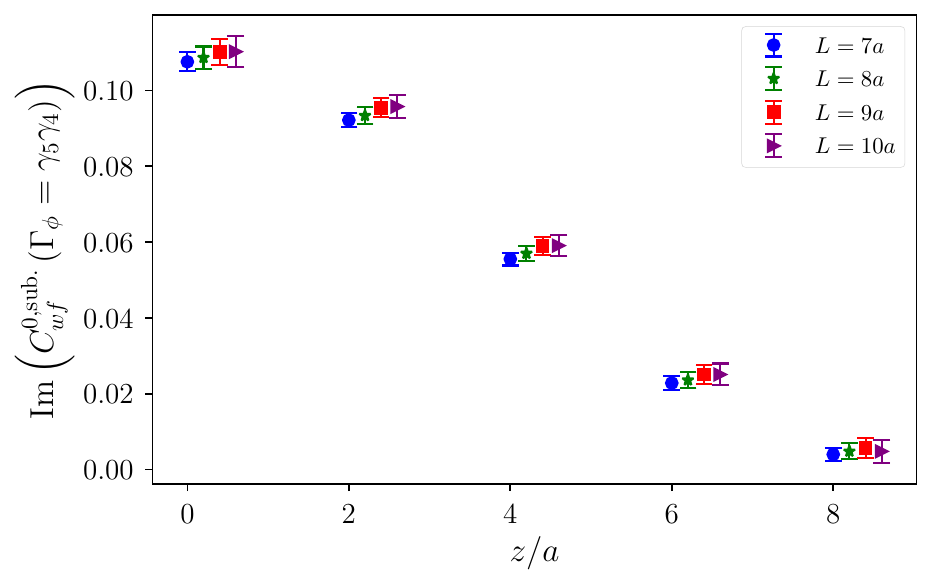}}
    \caption{Results on the real (left) and imaginary (right) parts of the subtracted correlation function ${C_{wf}^0}^{\rm sub.}$ employing the Dirac structure $\Gamma_{\phi} = \gamma_5\gamma_4$ using the cA211.53.24 ensemble and the valence pion mass $m_{\pi} = 830$~MeV. The selected boost is $P^z = 2.22$~GeV and the transverse separation $b = 2a$. We compare different $L$ values.}
    \label{fig:lindependence}
\end{figure}

In Fig.~\ref{fig:lindependence}, we show the ratio $C^{0,{\rm sub.}}_{wf} \equiv \tilde{\psi}^{\rm norm.}_{\Gamma_{\phi}}\left(b,z,L,P^z\right)/\sqrt{Z_E\left(2 L, b\right)}$, which, for sufficiently large values of $L$ yields the subtracted quasi-TMD~WF $\psi^{\rm sub}_{\Gamma_{\phi}}\left(b,z,P^z\right)$. The ratio is shown at $b=2a$ as a function of $z$, for four values of the staple parameter $L$ in the range $[7a,10a]$, using the cA211.53.24 ensemble and the valence pion mass $m_{\pi} = 830$~MeV. As can be seen, both the real and imaginary parts exhibit no significant dependence on $L$ within this range. The same behavior is observed across all other values of the parameter $b$, pion masses, Dirac structures, and ensembles considered in this work. Therefore, in the remainder of our analysis, we take the value of $C^{0,{\rm sub.}}_{wf}$ at $L=8a$ as representative of the subtracted quasi-TMD~WF $\psi^{\rm sub}_{\Gamma_{\phi}}\left(b,z,P^z\right)$, corresponding to a physical staple length of $0.744$ fm. 
\begin{figure}[hp!]
    \centering
    \subfigure{\includegraphics[width=0.49\linewidth]{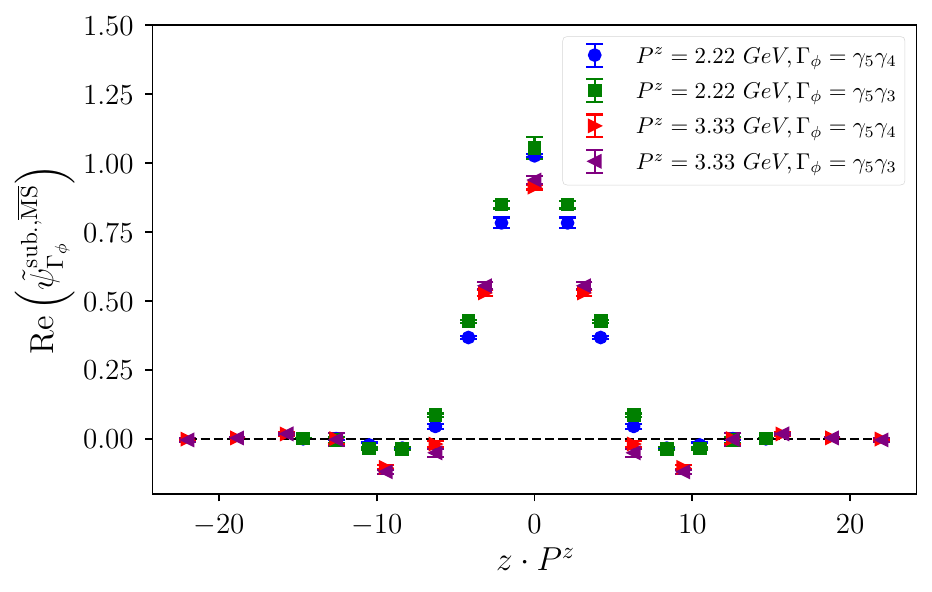}}%
    \subfigure{\includegraphics[width=0.49\linewidth]{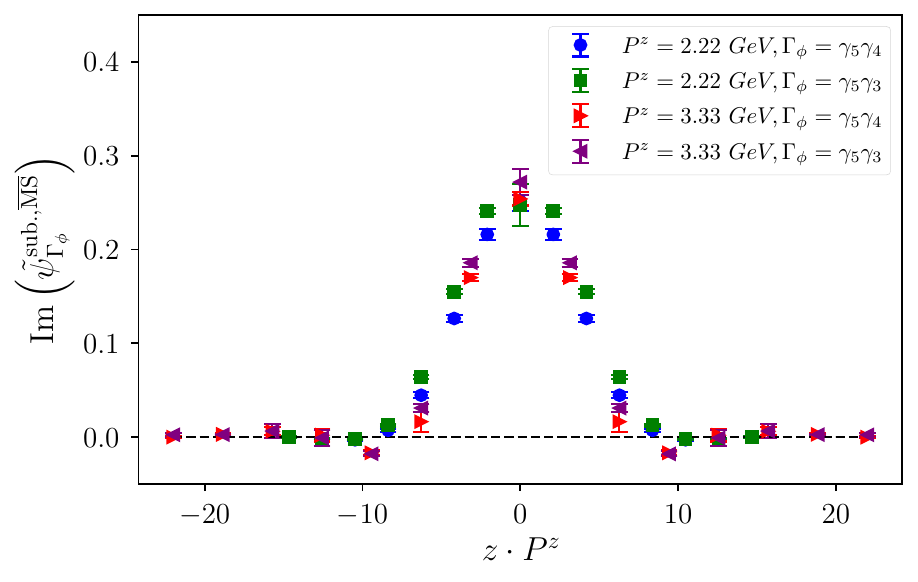}}\\
    \subfigure{\includegraphics[width=0.49\linewidth]{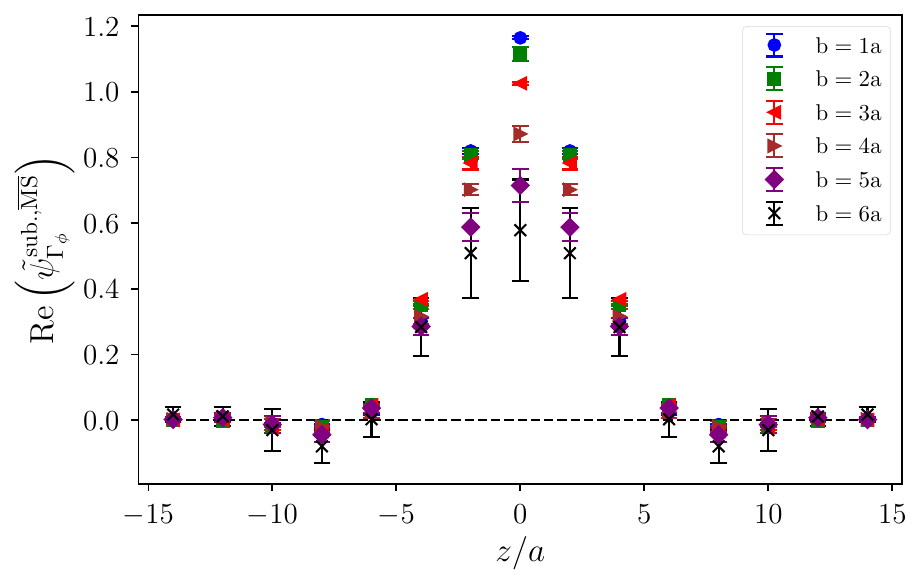}}
    \subfigure{\includegraphics[width=0.49\linewidth]{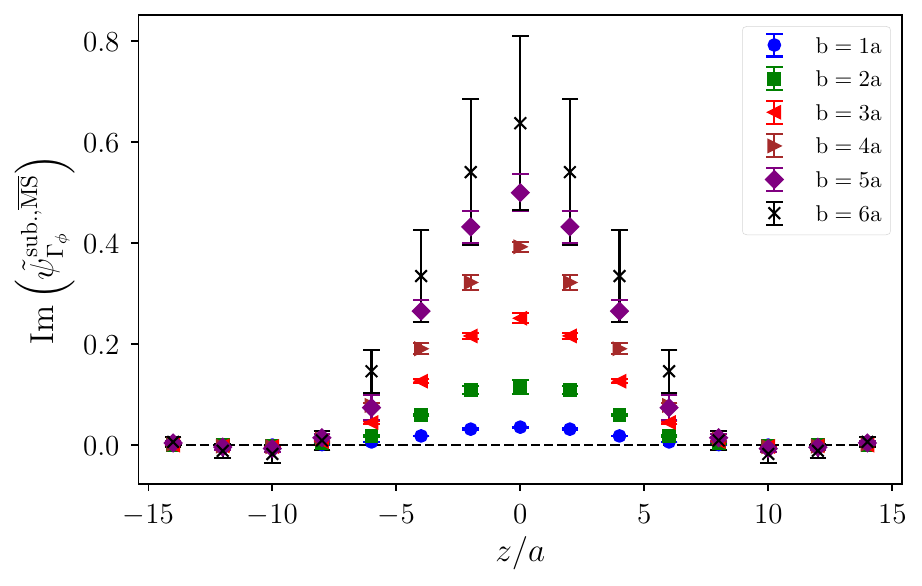}}
    
    \caption{Results for the real (left) and imaginary (right) parts of the $\overline{\rm MS}$-renormalized subtracted quasi-TMD~WF $\tilde{\psi}_{\Gamma_{\phi}}^{{\rm sub.},\overline{\rm MS}}$ at 2~GeV, using the cA211.53.24 ensemble and the valence pion mass $m_{\pi} = 830$~MeV. Top: Results for the leading-twist Dirac structures $\Gamma_{\phi} = \gamma_5\gamma_4$ and $\Gamma_{\phi} = \gamma_5\gamma_3$ at $b=3a$, $P^z = 2.22$~GeV or $P^z = 3.33$~GeV. Bottom: Results for $\Gamma_{\phi} = \gamma_5\gamma_4$ at $P^z = 2.22$~GeV and for different $b$ values. }
    \label{fig:psi_coordinate}
\end{figure}

After employing the SDR renormalization procedure described in Section~\ref{sec:renormalization}, we show in Fig.~\ref{fig:psi_coordinate}, results for the real and imaginary parts of the $\overline{\rm MS}$-renormalized subtracted quasi-TMD~WF $\tilde{\psi}^{{\rm sub.},\overline{\rm MS}}_{\Gamma_\phi} \left(z, b, P^z \right)$ at the reference scale of 2~GeV, for the Dirac structures $\Gamma_{\phi} = \gamma_5 \gamma_3$ and $\Gamma_{\phi} = \gamma_5 \gamma_4$. The plots in the top panel correspond to $b = 3a$ and boosts $P^z = 2.22$~GeV and $P^z=3.33$~GeV. The plots in the bottom panel present results of $\Gamma_{\phi} = \gamma_5 \gamma_4$ for a range of $b$ values at $P^z= 2.22$~GeV. The results are obtained for the cA211.53.24 ensemble and using the valence pion mass $m_{\pi} = 830$~MeV. As the boost increases, the wave function exhibits a more rapid decay with respect to $z$. The faster decay is consistent with the expected behavior of the Collins–Soper kernel discussed in Section~\ref{sec:results_collins}. Furthermore, we observe that the real (imaginary) part decreases (increases) with the transverse separation $b$. As the momentum increases, the results from the two Dirac structures tend to converge. At the largest momentum, $P^z=3.33$ GeV, they agree well for both the real and imaginary parts. A discrepancy is observed at the smallest momentum, $P^z=2.22$ GeV, indicating the presence of momentum-dependent systematic effects.
\begin{figure}[hp!]
    \centering
     \subfigure{\includegraphics[width=0.49\linewidth]{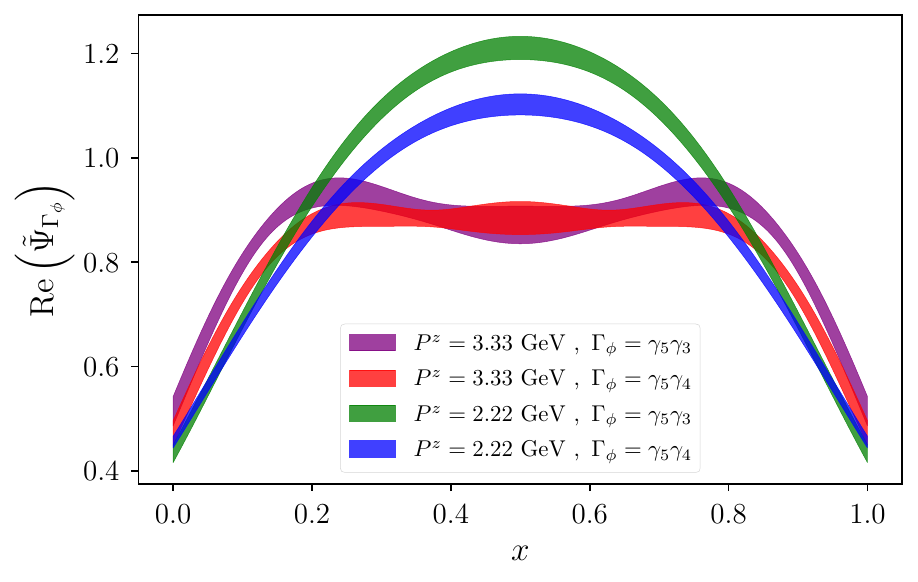}}%
     \subfigure{\includegraphics[width=0.49\linewidth]{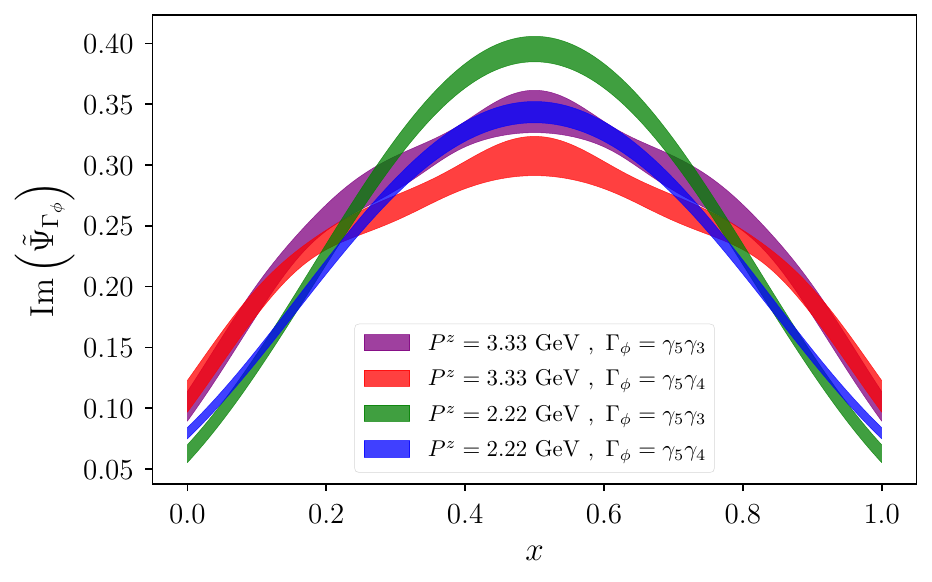}} \\
    \subfigure{\includegraphics[width=0.49\linewidth]{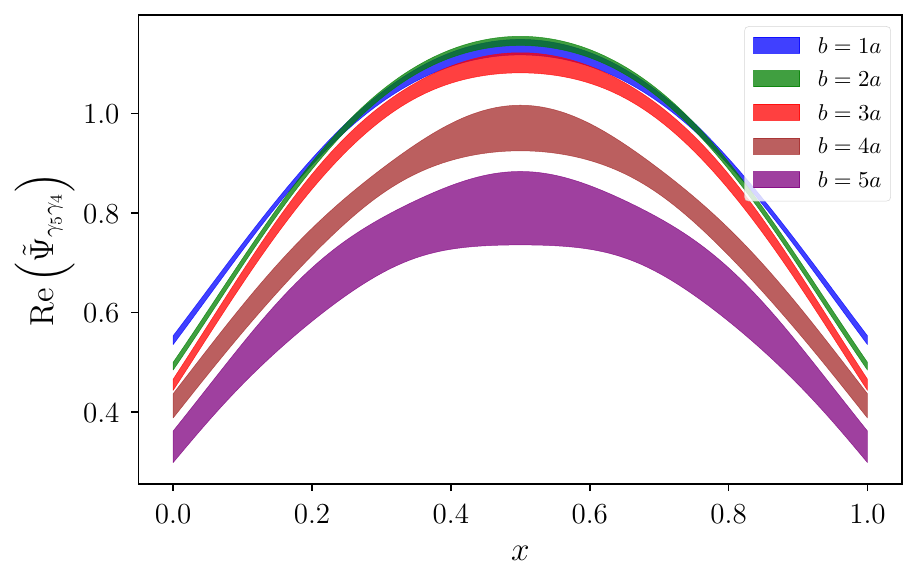}}
    \subfigure{\includegraphics[width=0.49\linewidth]{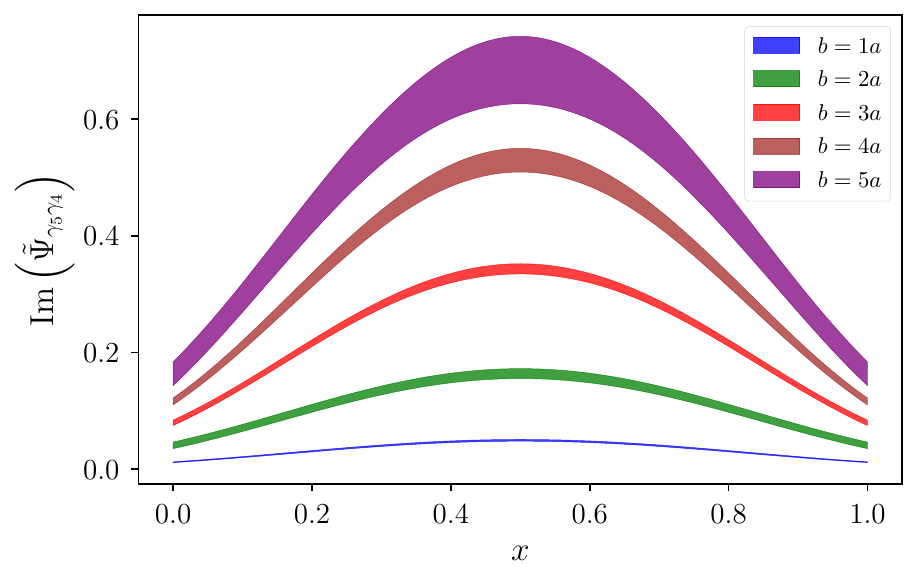}}
    
    \caption{Results for the real (left) and imaginary (right) parts of the momentum-space normalized and subtracted quasi-TMD~WF $\tilde{\Psi}_{\Gamma_{\phi}}$ in the $\overline{\rm MS}$ scheme at 2~GeV, using the cA211.53.24 ensemble and the valence pion mass $m_{\pi} = 830$~MeV. Top: Results for the leading-twist Dirac structures $\Gamma_{\phi} = \gamma_5\gamma_4$ and $\Gamma_{\phi} = \gamma_5\gamma_3$ at $b=3a$, $P^z = 2.22$~GeV or $P^z = 3.33$~GeV. Bottom: Results for $\Gamma_{\phi} = \gamma_5\gamma_4$ at $P^z = 2.22$~GeV and for different $b$ values.}
    \label{fig:psi_momentum}
\end{figure}

The momentum-space quasi-TMD~WF $\tilde{\Psi}$ is obtained from the coordinate-space values, by performing a discrete Fourier Transform (DFT) over the values of the staple parameter $z$ 
\begin{equation}
\tilde{\Psi}_{\Gamma_{\phi}}\left(x,b,P^z,\mu\right) = \sum_{-z_{max}}^{z_{max}} \frac{P^z}{2\pi}e^{-\mathit{i}(x-\frac{1}{2})zP^z}\tilde{\psi}^{{\rm sub.}}_{\Gamma_{\phi}}\left(z,b,P^z\right) ,
\end{equation}
where $z_{max}$ was set to $14a$ in our calculation. We verify that convergence is reached for this value of $z_{\rm max}$. 
Unlike the momentum-space quasi-PDF extraction, the TMD case exhibits better convergence to zero at large $z$, with smaller associated errors. This behavior improves the reliability of the DFT method in reconstructing quasi-TMD observables in momentum space. A key factor behind this is the better controlled $z$-dependent systematics in the TMD case. In particular, the linear divergences in TMD correlators are independent of $z$ for the staple geometry considered in this work (see Fig. \ref{fig:staple}). These divergences depend on the total length of the staple, given by $2L+b$. While the lengths of the individual segments vary with $z$ (specifically the two transverse parts), their total length remains independent of $z$. Additionally, TMD matrix elements decay faster with $z$ than their quasi-PDF counterparts, reducing the sensitivity to the large-$z$
region and allowing safe truncation of the DFT method at relatively small $z_{\rm max}$. In contrast, in the quasi-PDF case the linear divergences scale linearly with $z$. Although nonperturbative renormalization methods, such as RI$'$/MOM or ratio schemes, can be applied to cancel these divergences, residual $z$-dependent effects remain more pronounced in quasi-PDFs. As a result, the choice of $z_{\rm max}$ in the PDF case becomes critical, since statistical and systematic uncertainties grow rapidly at large $z$~\cite{Alexandrou:2019lfo}.

Results for the momentum-space quasi-TMD~WF are presented in Fig.~\ref{fig:psi_momentum} as a function of the momentum fraction $x$. As expected, we observe a non-negligible imaginary part, which becomes more pronounced with increasing $b$. However, as discussed in Section~\ref{sec:results_collins}, this imaginary contribution does not affect the final physical observables. We also observe that 
a plateau region appears for $0.3<x<0.7$ in both real and imaginary parts, especially for large values of $b$. A comparison between the two leading-twist Dirac structures, $\gamma_5\gamma_4$ and $\gamma_5\gamma_3$, shows good agreement. Based on this agreement, and for simplicity, we focus in the remainder of this work on the Dirac structure $\gamma_5\gamma_4$, while results obtained with $\gamma_5\gamma_3$ are expected to be comparable.

\subsection{Collins-Soper Kernel}
\label{sec:results_collins}

\begin{figure}[!ht]
    \centering
    \subfigure{\includegraphics[width=0.49\linewidth]{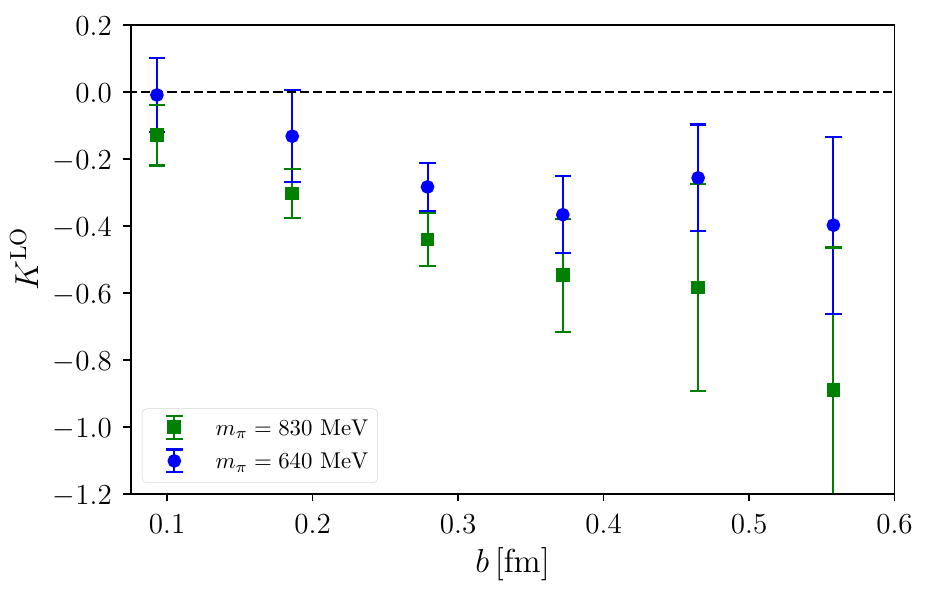}}
    \subfigure{\includegraphics[width=0.49\linewidth]{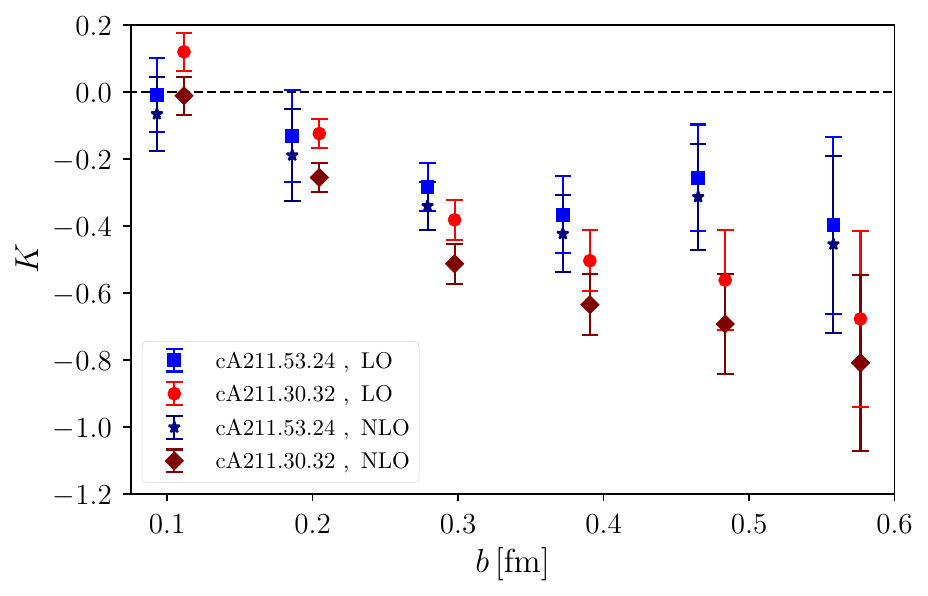}}

    \caption{ Results for the Collins-Soper kernel $K$. Left: The LO contributions are compared for $m_\pi=640$~MeV and $m_\pi=830$~MeV using the cA211.53.24 ensemble and boosts $P^z=2.22$ and $3.33$~GeV. Right: LO and NLO contributions are compared for the ensembles cA211.53.24 (using boosts $P^z=2.22$ and $3.33$~GeV) and cA211.30.32 (using boosts $P^z=1.63$ and $2.45$~GeV) at valence pion mass $m_{\pi} = 640$~MeV. the cA211.30.32 ensemble results are shifted to the right for better readability.}
    \label{fig:CSK}
\end{figure}

As described in Section~\ref{sec:CSKernel}, to extract the Collins-Soper kernel we take the ratio of momentum-space quasi-TMD~WFs 
at two different large momenta. 
\begin{figure}[!hpt]
    \centering
    \subfigure{\includegraphics[width=0.49\linewidth]{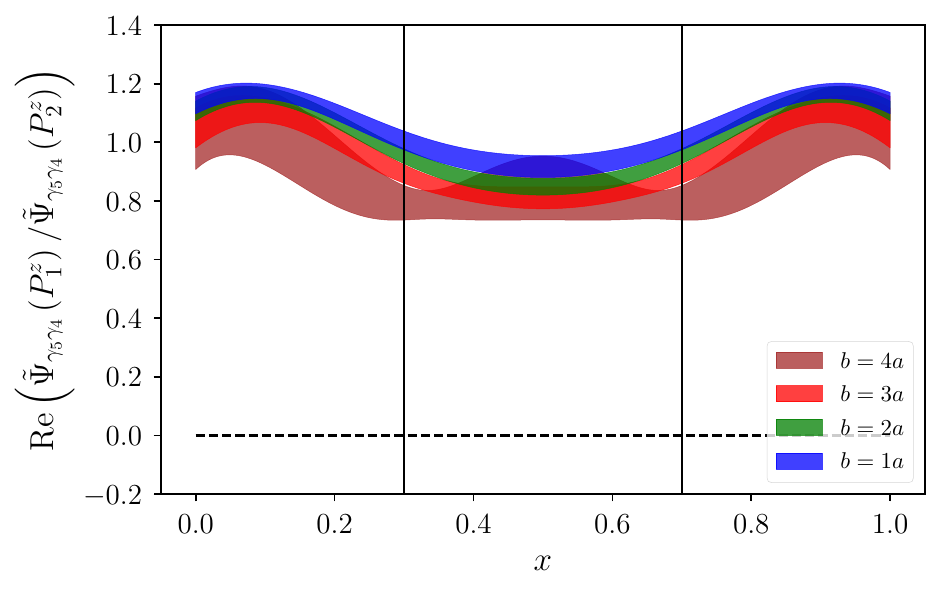}}%
    \subfigure{\includegraphics[width=0.49\linewidth]{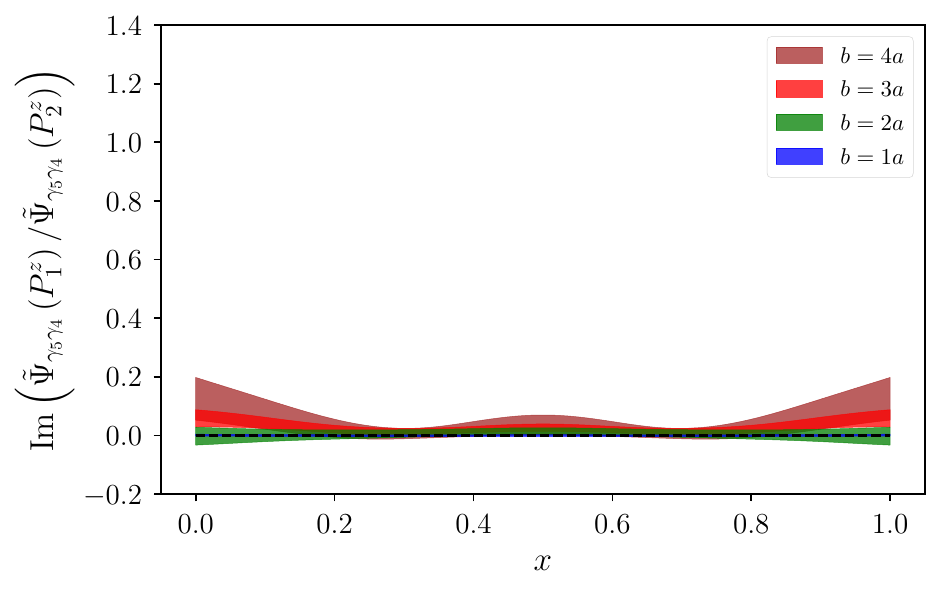}}

    \caption{ Results for the real (left) and imaginary (right) parts of the ratio of momentum-space quasi-TMD~WFs $\tilde{\Psi}_{\gamma_5\gamma_4}$ for boosts $P_1^z = 3.33$~GeV and $P_2^z = 2.22$~GeV and for different values of $b$. The results are shown for the cA211.53.24 ensemble and with valence pion mass $m_{\pi} = 830$~MeV. The vertical lines on the left panel show the region in which the ratio is approximately constant.}
    \label{fig:ratio}
\end{figure}
In  Fig.~\ref{fig:ratio}, 
we show the real and imaginary parts of this ratio as a function of the momentum fraction $x$, for the Dirac structure
$\gamma_5\gamma_4$ using boosts $P^z = 2.22$~GeV and $P^z = 3.33$~GeV on the cA211.53.24 ensemble and with valence pion mass $m_{\pi} = 830$~MeV.
In the range of $x \in \left[0.3 , 0.7\right]$, the ratio is approximately constant, consistent with the expected $x$-independence of the Collins-Soper kernel.
Closer to the endpoints $x=0$ and $x=1$, the
results are affected by power corrections of the form $M/(xP^z)^2$ and $M/((1-x)P^z)^2$, respectively.
Although the imaginary part should be zero, since the Collins-Soper kernel is a real quantity, we observe a small but nonzero imaginary contribution. 
However, this contribution is significantly smaller than the real part and becomes consistent with zero at larger values of $b$. Therefore, in the following analysis, we neglect the imaginary part. Results are presented up to $b=4a$ ; beyond this point, statistical uncertainties increase substantially, and higher statistics would be required.
For each $b$, we perform a weighted average of the ratio over the range $x \in \left[0.3 , 0.7\right]$. The resulting values are used to extract the Collins-Soper kernel via Eq.~\eqref{eq:kernel}. In the left plot of Fig.~\ref{fig:CSK}, we show the leading order (LO) result for the Collins-Soper kernel $K^{LO}$, where the $\mathcal{O} (\alpha_s)$ term of Eq.~\eqref{eq:kernel} is omitted. We investigate the dependence of $K^{LO}$ on the pion mass by using two valence masses, $m_\pi=640$~MeV and $m_\pi=830$~MeV, on the cA211.53.24 ensemble. For the lighter pion mass, we employ double the statistics for the smaller boost in order to achieve uncertainties comparable to those of the heavier mass. We find that the results using the two different masses are in agreement. 
\begin{figure}[!h]
    \centering
     \subfigure{\includegraphics[width=0.6\linewidth]{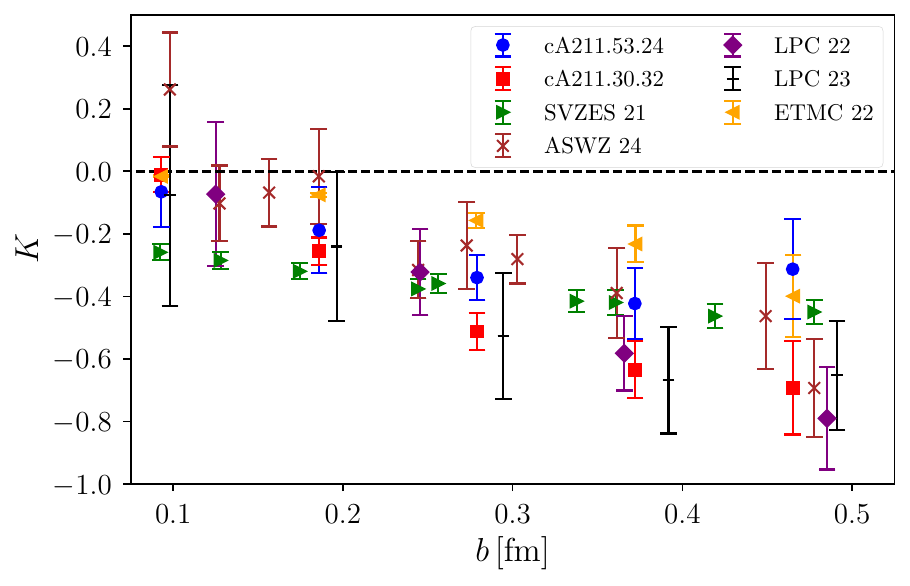}}%

    \caption{Comparison of our results for the Collins-Soper kernel other lattice QCD results. Our results are obtained using  the cA211.53.24 (blue circles) and cA211.30.32 (red squares) ensembles, at pion mass $m_{\pi} = 640$~MeV, using ratios of quasi-TMD WFs at  boosts of 2.22~GeV and 3.33~GeV for cA211.53.24 and 1.63 and 2.45~GeV for cA211.30.32. Other results are obtained from Refs.~\cite{Li:2021wvl} (ETMC 22), ~\cite{Schlemmer:2021aij} (SVZES 21),~\cite{LatticePartonLPC:2022eev} (LPC 22),~\cite{LatticePartonLPC:2023pdv} (LPC 23), and ~\cite{Avkhadiev:2024mgd} (ASWZ 24).}
    \label{fig:CSK_comparison}
\end{figure}
In the right plot of Fig.~\ref{fig:CSK}, we compare LO and next-to-leading order (NLO) $\mathcal{O}\left(\alpha_s\right)$  results for the Collins-Soper kernel using  the cA211.53.24 and cA211.30.32 ensembles, both at a fixed pion mass of $m_{\pi} = 640$~MeV. For cA211.53.24 we use boosts $P^z = 2.22$~GeV and $3.33$~GeV, while for cA211.30.32, the boosts are $P^z = 1.63$~GeV and $2.45$~GeV. For cA211.53.24, the difference between LO and NLO results is smaller than the statistical uncertainties and is therefore considered negligible. For cA211.30.32, the difference is more pronounced, primarily due to the use of smaller boost momenta leading to larger NLO corrections (see Eq.~\eqref{eq:kernel}). 
Comparing the two ensembles, the results for $K$ are compatible between each other within two standard deviations, suggesting that finite volume effects are under control. 

In Fig.~\ref{fig:CSK_comparison}, we compare our results for the Collins-Soper kernel with our previous lattice calculation~\cite{Li:2021wvl} and with other collaborations
~\cite{Schlemmer:2021aij,LatticePartonLPC:2022eev,LatticePartonLPC:2023pdv,Avkhadiev:2024mgd}. The extraction method varies across these studies. In SVZES 21~\cite{Schlemmer:2021aij}, the Collins-Soper kernel is obtained from ratios of the first Mellin moments of quasi-TMD~PDFs. In contrast, all other works including ours extract the kernel from ratios of quasi-TMD~WFs. The Lattice Parton Collaboration, LPC 22~\cite{LatticePartonLPC:2022eev} and LPC 23~\cite{LatticePartonLPC:2023pdv} use a setup similar to ours, while ASWZ 24~\cite{LatticePartonLPC:2023pdv} differs by adopting a modified normalization of the quasi-TMD WFs, accounting for operator mixing, and performing a continuum extrapolation based on results from three lattice spacings. In our previous study, ETMC 22~\cite{Li:2021wvl}, the Collins-Soper kernel was extracted only at leading order, where ratios were computed by integrating the momentum-space quasi-TMD WFs over the momentum fraction $x$, resulting in ratios of position-space quasi-TMD~WFs at $z=0$, namely using
 \begin{equation}
     \frac{\int dx \tilde{\Psi}^\pm_{\Gamma_{\phi}} (x,b,\mu,\zeta_{z1})}{\int dx \tilde{\Psi}^\pm_{\Gamma_{\phi}} (x,b,\mu,\zeta_{z2})} = \frac{\tilde{\psi}^{{\rm sub.},\,\pm}_{\Gamma_{\phi}}\left(b,0,P^z_1\right)}{\tilde{\psi}^{{\rm sub.},\,\pm}_{\Gamma_{\phi}}\left(b,0,P^z_2\right)} = \frac{\tilde{\psi}^{\rm norm.}_{\Gamma_{\phi}}\left(b,0,\pm L, P^z_1\right)}{\tilde{\psi}^{\rm norm.}_{\Gamma_{\phi}}\left(b,0,\pm L, P^z_2\right)}.
 \end{equation}
 This approach avoids the direct computation of the Fourier transform of the quasi-TMD WFs. This explains the smaller statistical uncertainties of our older study compared to our current analysis. However, NLO corrections, integrated over $x$, are missing. From Fig.~\ref{fig:CSK_comparison}, we conclude that our current results are consistent with the more recent results of ASWZ 24, LPC 22 and LPC 23.

\subsection{Reduced Soft Function}
\label{sec:results_soft}

As described in Section~\ref{sec:soft}, to extract the reduced soft function, we calculate the normalized form factor $F^{\rm norm}_\Gamma$, in addition to the momentum-space quasi-TMD~WF computed in Section~\ref{sec:results_tmdwf} (see Eq.~\eqref{eq:soft}). 

\begin{figure}[!h]
    \centering
        \subfigure{\includegraphics[width=0.49\linewidth]{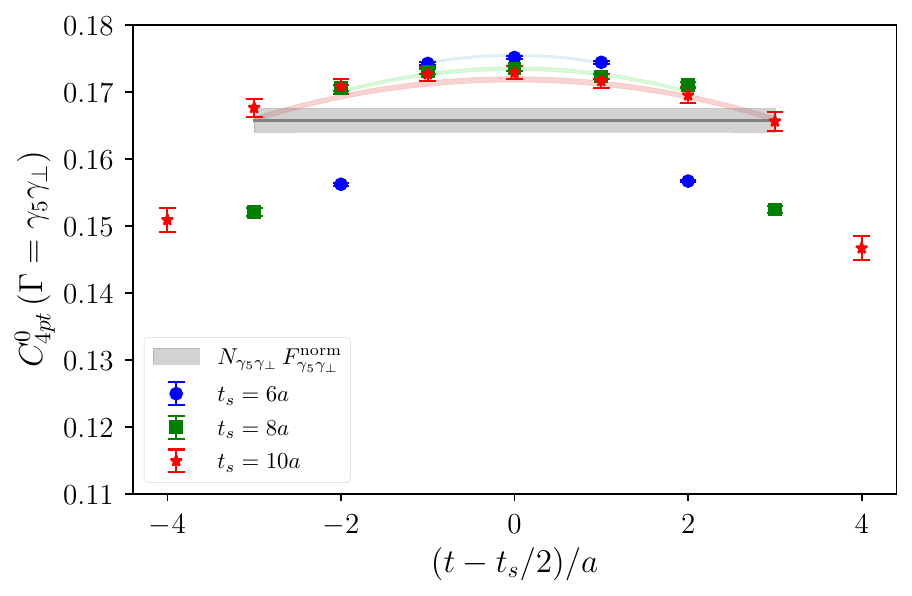}}%
    \subfigure{\includegraphics[width=0.49\linewidth]{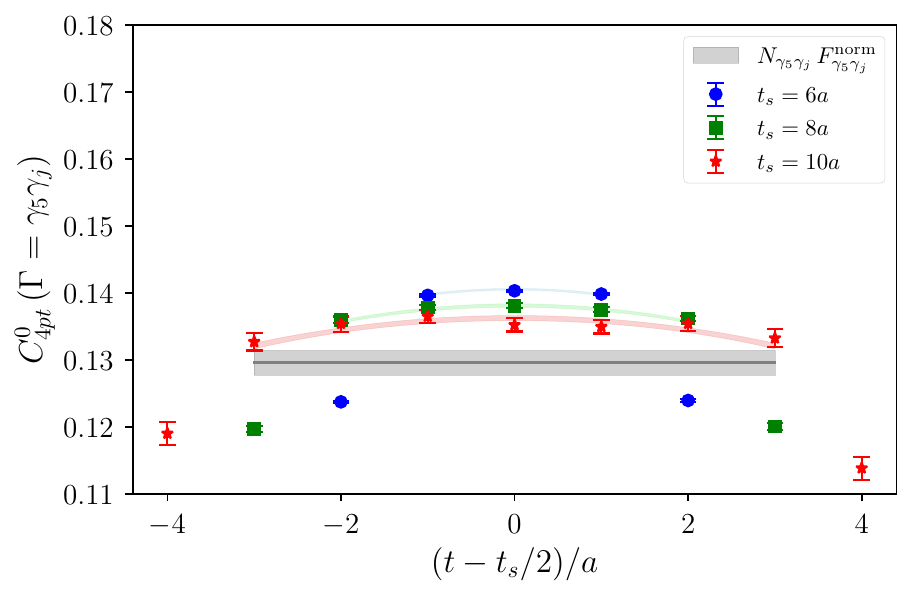}}\\
    \subfigure{\includegraphics[width=0.49\linewidth]{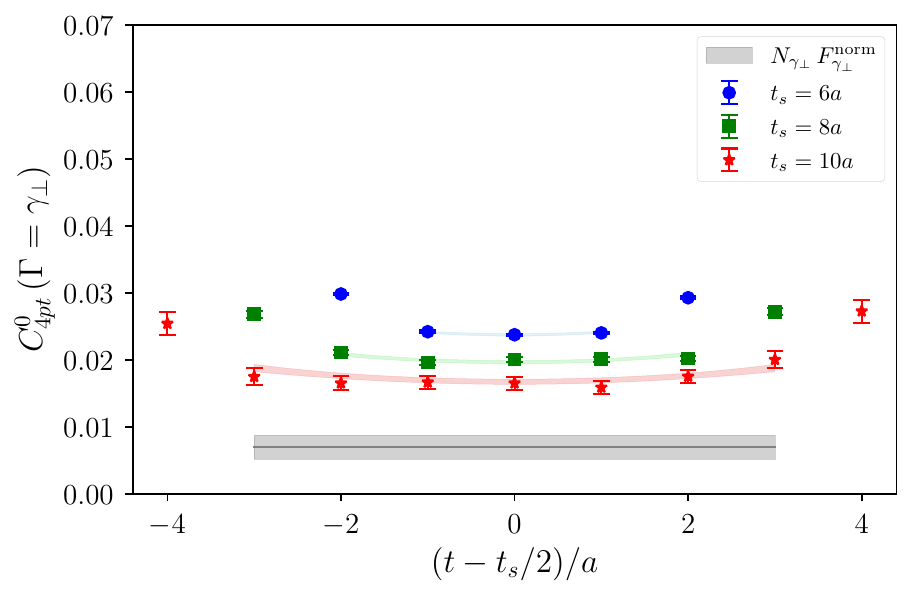}}%
    \subfigure{\includegraphics[width=0.49\linewidth]{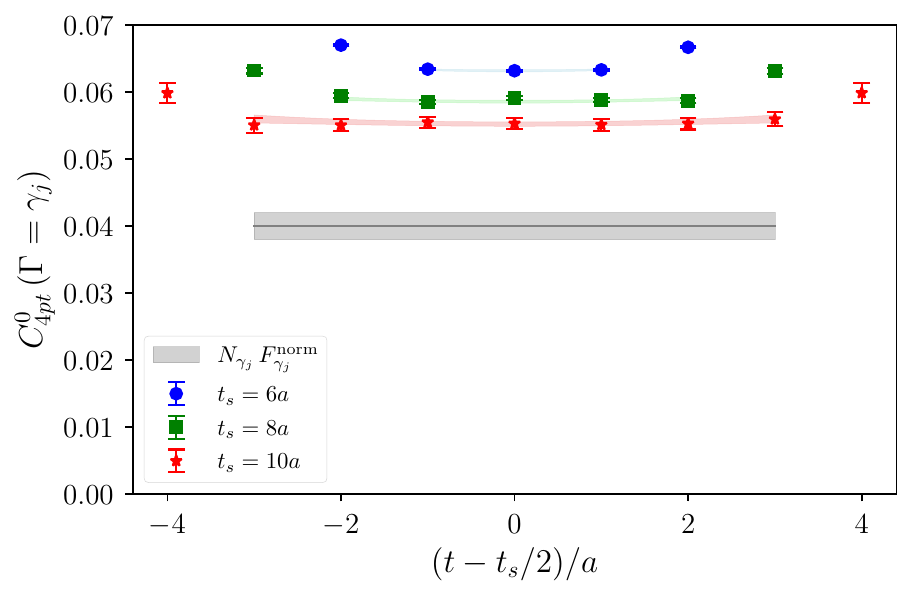}}
    \caption{Results of the normalized four-point correlation function $C_{4pt}^0$ at $P^z = 2.22$~GeV, $b=2a$ using the cA211.53.24 ensemble and the valence pion mass $m_{\pi} = 830$~MeV, for  $\Gamma = \gamma_5\gamma_{\perp}$ (top left), $\Gamma = \gamma_5\gamma_j$ (top right), $\Gamma = \gamma_{\perp}$ (bottom left) and $\Gamma = \gamma_j$ (bottom right). Data for source-sink time separation $t_s=6a,\, 8a$ and $10a$ are shown, respectively, in blue, green and red points. The gray bands represent the joint fit extraction of the meson form factor ground state contribution $N_\Gamma \, F_\Gamma^{\rm norm}(b,P^z)$.
    }
    \label{fig:fourp}
\end{figure}
\begin{figure}[!h]
    \centering
    \subfigure{\includegraphics[width=0.49\linewidth]{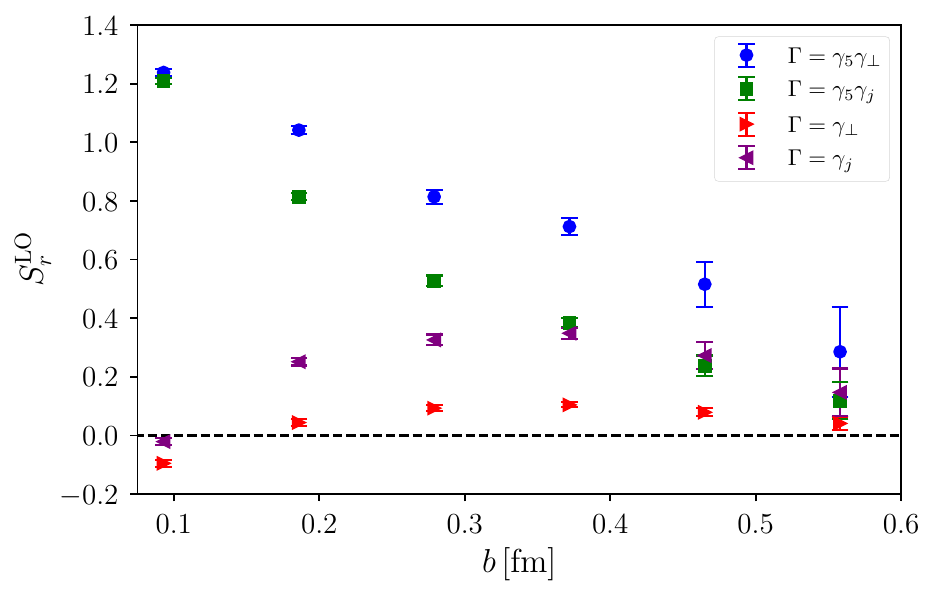}}%
    \subfigure{\includegraphics[width=0.49\linewidth]{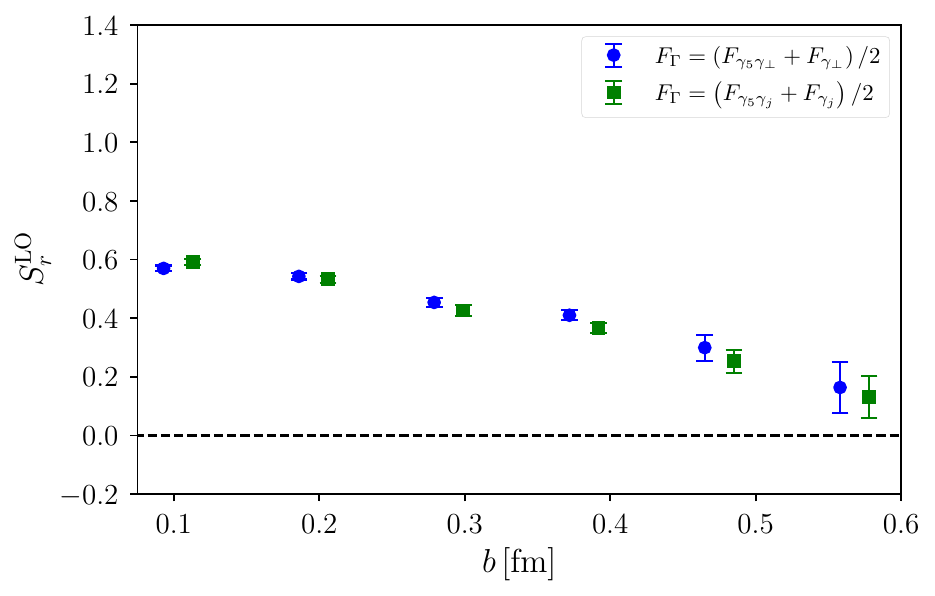}} \\
    \subfigure{\includegraphics[width=0.49\linewidth]{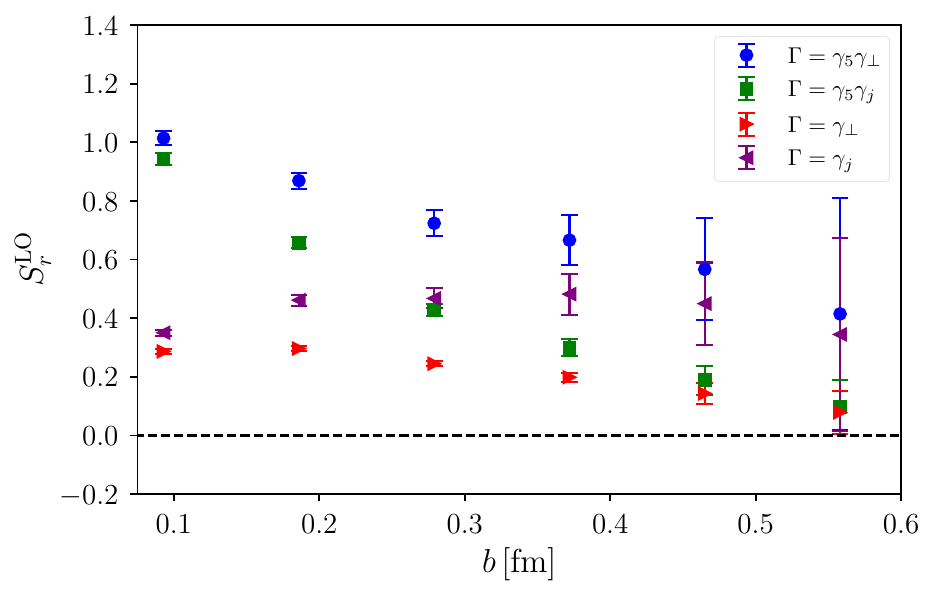}}%
    \subfigure{\includegraphics[width=0.49\linewidth]{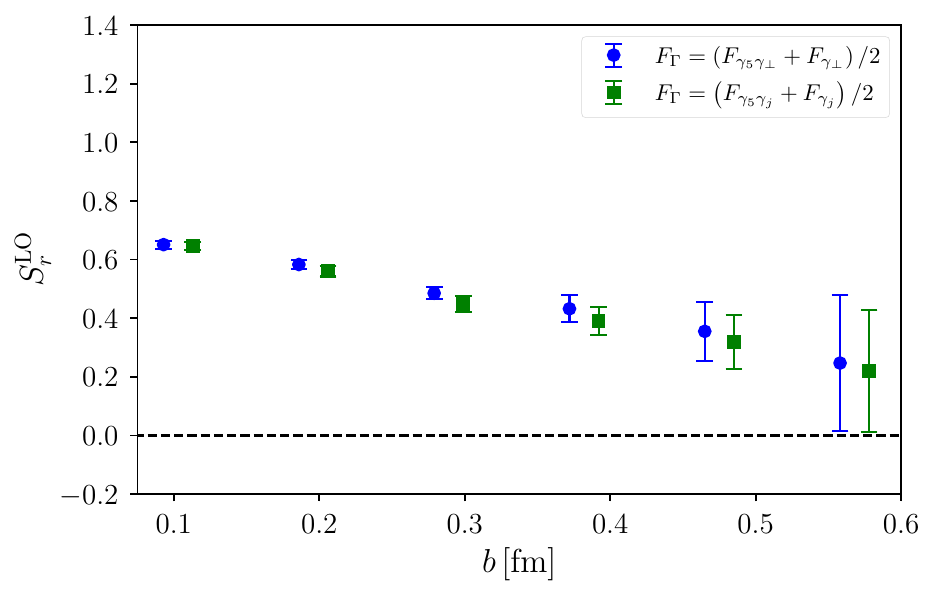}} 
    \caption{Results for the reduced soft function $S_r$ at LO using the cA211.53.24  and the valence pion mass $m_{\pi} = 830$~MeV. Top panel shows results by using the Dirac structures $\gamma_{\perp}$, $\gamma_j$, $\gamma_5\gamma_{\perp}$ and $\gamma_5\gamma_{j}$ (left) and the combinations $(\gamma_{\perp}+ \gamma_5\gamma_{\perp})/2$, and $(\gamma_j+\gamma_5\gamma_j)/2$ (right) for the boost momentum $P^z = 2.22 $~GeV. Bottom panel shows the corresponding results for the boost momentum $P^z = 3.33$~GeV. The $F_{\Gamma} = (\gamma_j+\gamma_5\gamma_j)/2$ results on the right column have been shifted for better readability.}
    \label{fig:soft}
\end{figure}
\begin{figure}[h!]
    \subfigure{\includegraphics[width=0.49\linewidth]{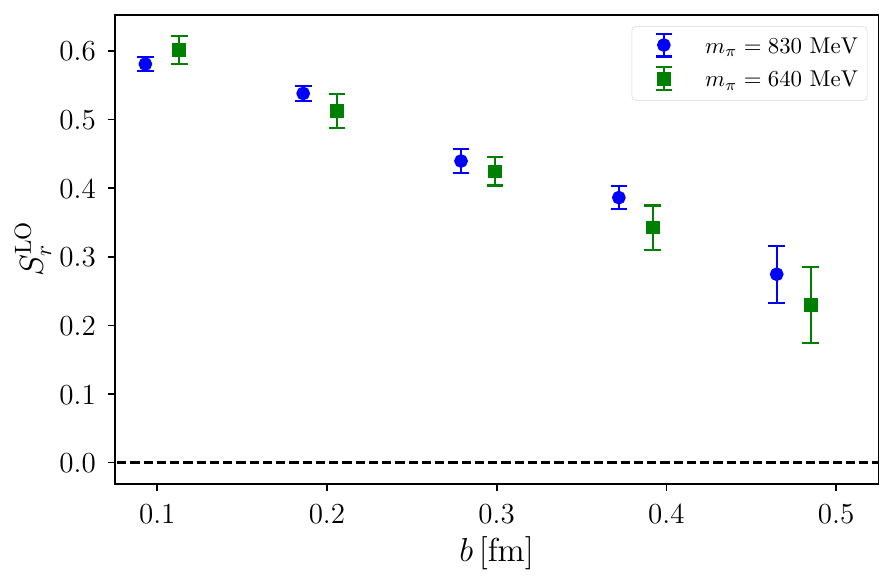}}   
    \subfigure{\includegraphics[width=0.49\linewidth]{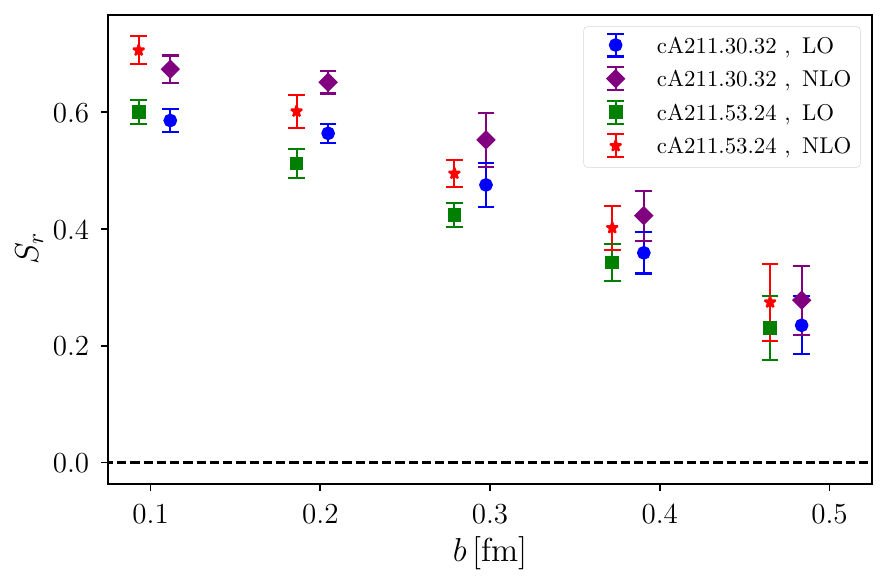}}
    
    \caption{Left: Results for the reduced soft function $S_r^{LO}$ at pion masses 640~MeV and 830~MeV using the ensemble cA211.53.24 and boost $P^z = 2.22$~GeV. Right: Results for the reduced soft function at LO and NLO at the pion mass $m_{\pi}=640$~MeV, using the cA211.53.24 and cA211.30.32 ensembles with boosts $P^z = 2.22$~GeV and $P^z = 2.45$~GeV, respectively. The cA211.30.32 ensemble results have been shifted for better readability.}
    \label{fig:soft2}
\end{figure}
In Fig.~\ref{fig:fourp}, we present the normalized four-point function $C_{4pt}^0$ at fixed transverse separation $b=2a$, plotted as a function of $(t-t_s/2)/a$ for three source-sink time separations: $t_s = 6a$, $8a$, and $10a$. We consider the Dirac structures $\Gamma = \gamma_5 \gamma_{\perp}, \gamma_5 \gamma_{j}, \gamma_{\perp}$ and $\gamma_{j}$, where $\gamma_{\perp}$ refers to the gamma matrix in the direction of the transverse separation, and $\gamma_{j}$ corresponds to the direction orthogonal to both the boost and the transverse separation.
We extract the ground-state contribution of $C_{4pt}^0$ using  two-state fits to all three values of $t_s$ and for each value of $b$.

Although the reduced soft function is expected to be independent of the Dirac structure used in the four-point function, we observe  in  Fig.\ref{fig:soft} notable differences when using different Dirac structures, for both  momentum boosts, where  the reduced soft function is computed using $\mathcal{H}_{F_\Gamma}$  at  LO, i.e., $|J_1|^2$ (see Eq.~\eqref{eq:final_denominator}), denoted as $S_r^{LO}$. The dependence on the Dirac structure was understood in our previous work~\cite{Li:2021wvl}, where each structure was shown to have different higher-twist contaminations. As we observe, the contaminations become less significant as the momentum boost increases from 2.22~GeV to 3.33~GeV. To suppress higher-twist effects, we construct appropriate combinations of four-point functions involving different Dirac structures. This approach is motivated by the LO factorization of the meson form factor~\cite{Li:2021wvl}:
\begin{equation}
F_{\Gamma}(P^z,b) = S_r(b) \, N_{\Gamma} \sum_{\Gamma' = \gamma_5 \gamma_4, \gamma_5 \gamma_3} |{\tilde{\psi}_{\Gamma'}^{{\rm sub.},\pm}}(b,0,P^z)|^2 + \sum_{\Gamma^{'} \ne \gamma_5\gamma_4,\gamma_5\gamma_3}
{(S_r)}_{\Gamma'}(b) \ N_{\Gamma \Gamma '} \  |{\tilde{\psi}_{\Gamma'}^{{\rm sub.},\pm}}(b,0,P^z)|^2+ \mathcal{O} (\alpha_s),
\label{eq:factorization}
\end{equation}
where the first term isolates leading-twist contributions, and the second term contains higher-twist contaminations. The coefficients $N_{\Gamma \Gamma'}$ arise from Fierz rearrangement, expressed as: 
\begin{equation}
\bar{u} (b) \Gamma u (b) \bar{d} (0) \Gamma d(0) = \sum_{\Gamma'} N_{\Gamma \Gamma'} \bar{u}(b)\Gamma' d (0) \bar{d} (0) \Gamma' u(b) ,
\end{equation}
with $N_{\Gamma \Gamma'} = \frac{1}{16N_C} {\rm Tr}\left(\Gamma\Gamma'\Gamma\Gamma'\right)$.
By taking the following linear combinations of four-point functions~\cite{Li:2021wvl}, we eliminate the higher-twist contributions (i.e., the second term in Eq.~\eqref{eq:factorization}):
\begin{equation}
\frac{\left(F_{\gamma_{\perp}}+F_{\gamma_5\gamma_{\perp}}\right)}{2} \quad , \quad \frac{\left(F_{\gamma_j}+F_{\gamma_5\gamma_j}\right)}{2}.
\label{eq:combinations}
\end{equation}
These combinations isolate the leading-twist behavior and allow for a cleaner extraction of the reduced soft function across different boost momenta.
In the right panel of Fig.~\ref{fig:soft}, we present $S_r^{LO}$ obtained from the combinations defined in Eq.\eqref{eq:combinations}. The results from the two combinations are in good agreement. In what follows, we take their average.
In the left plot of Fig.~\ref{fig:soft2}, we compare $S_r^{LO}$, averaged over the Dirac structure combinations discussed above, for two valence pion masses, $m_{\pi} = 640$~MeV and $m_{\pi} = 830$~MeV, using the cA211.53.24 ensemble at a fixed boost momentum $P^z = 2.22$~GeV. The results are compatible. 
In the right plot of Fig.~\ref{fig:soft2}, we compare results for $S_r^{LO}$, with those  obtained by including NLO corrections in $\mathcal{H}_{F_\Gamma}$ (see Eq.~\eqref{eq:final_denominator}), denoted as $S_r^{NLO}$. The comparison is performed for the two ensembles under study at a pion mass of $m_\pi = 640$~MeV, using the following momentum boosts: $P^z = 2.22$~GeV for the cA211.53.24 ensemble and $P^z = 2.45$~GeV for the cA211.30.32 ensemble. The results from the two ensembles agree within one standard deviation, indicating that there are no significant finite-volume effects. However, noticeable differences are observed between $S_r^{LO}$ and $S_r^{NLO}$, highlighting the impact of NLO corrections. In particular, the inclusion of NLO terms leads to a significant increase of $\sim 15–30 \%$ in the value of the soft function, depending on $b$. 
\begin{figure}[!ht]
    \centering
     \includegraphics[width=0.6\linewidth]{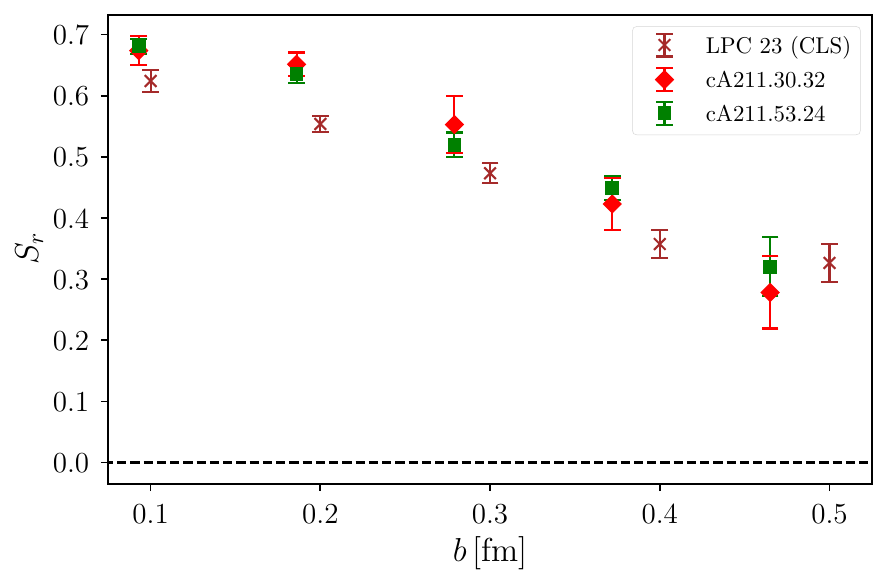}
    \caption{Comparison of our results for   the reduced soft function using ensembles cA211.53.24 and cA211.30.32 with the results from LPC~\cite{LatticePartonLPC:2023pdv}  using CLS configurations (LPC 23 (CLS)).}
    \label{fig:soft_comparison}
\end{figure}
In Fig.~\ref{fig:soft_comparison}, we compare our final results for the reduced soft function, including NLO corrections, with the lattice QCD results presented by LPC~\cite{LatticePartonLPC:2023pdv}. We do not include a comparison with our earlier results from Ref.~\cite{Li:2021wvl}, as that work was limited to LO accuracy and did not incorporate the matching factor for the quasi-TMD wave function to the $\overline{\rm MS}$ scheme, which was not yet available at the time of publication. 

\section{Conclusions}
\label{sec:conclusions}

In this work, we present a lattice QCD calculation of the Collins–Soper kernel and the reduced soft function, both of which are essential ingredients for the determination of transverse momentum-dependent parton distribution functions (TMD PDFs) from first principles. Our approach is based on the evaluation of quasi-TMD wave functions, which are computed using spatial correlation functions involving nonlocal quark bilinear operators connected by asymmetric staple-shaped Wilson lines.

The calculation is performed using two ensembles of $N_f = 2+1+1$ twisted-mass fermions with lattice spacing 0.093~fm, valence pion masses of 830~MeV and 640~MeV, and lattice  volumes of $24^3 \times 48$ and $32^3 \times 64$. The quasi-TMD WFs are analyzed using one-state and two-state fits to reliably isolate ground-state contributions. For the reduced soft function, we also compute four-point correlation functions involving bilocal quark bilinears and extract the corresponding meson form factors from their ground-state matrix elements.

Renormalization is carried out nonperturbatively using the Short Distance Ratio (SDR) scheme. Prior to this, we verify that operator mixing, predicted by symmetry arguments, is numerically negligible for the operators under study. After renormalization, we convert our results to the $\overline{\rm MS}$ scheme to facilitate comparison with phenomenological analyses.

The Collins–Soper kernel is extracted from ratios of momentum-space quasi-TMD WFs evaluated at different hadron momenta, while the reduced soft function is obtained via the factorization of the meson form factor in terms of quasi-TMD WFs. NLO corrections are incorporated in the extraction of both the Collins–Soper kernel and the reduced soft function, ensuring a consistent determination of the TMD~PDF matching at NLO. Their inclusion has a notable effect on both quantities and contributes to a reduction in systematic uncertainties. Our results are in good agreement with previous lattice studies using different discretizations and setups. 
This consistency supports the validity and robustness of the methodology used. 
Having both the Collins-Soper kernel and the reduced soft function, we  have all the necessary ingredients to compute physical TMD PDFs, over a range of transverse separations up to approximately $b = 0.36$ fm. 
 In future work, we plan to extend this analysis to ensembles at the physical pion mass and  extract the TMD PDFs for the proton, pion and kaon.

\begin{acknowledgments}

    C.A., S.B., G.S. and J.T. acknowledge  funding from the projects 
    "Unraveling the 3D Parton structure of the nucleon with lattice QCD (3D-nucleon)” 
    (contract id EXCELLENCE/0421/0043), NiceQuarks (contract id EXCELLENCE/0421/0195), Image-N (contract id EXCELLENCE/0524/0459) and Baryon8 (contract id POSTDOC/0524/0001) co-financed by the European Regional 
    Development Fund and the Republic of Cyprus through the Research and 
    Innovation Foundation. C.A. acknowledges funding from the European Union's research and innovation programme under the Marie Sklodowska-Curie Doctoral Networks action and Grant Agreement No. 101072344 and the projects 3D-N-LQCD and PDFs-LQCD funded by the University of Cyprus. 
	A.S and F.S are
funded in part by the Deutsche Forschungsgemeinschaft (DFG, German Research Foundation) as part of the CRC
1639 NuMeriQS – project No. 511713970. 
    K.C. is supported by the National 
    Science Centre (Poland) OPUS grant No. 2021/43/B/ST2/00497. 
    M.C. acknowledges financial support by the U.S. Department of Energy, Office of Nuclear Physics,  under Grant No.\ DE-SC0025218.
    
    This work used computational resources from the John von Neumann-Institute 
    for Computing on the Juwels booster system at the research center in Juelich, 
    under the project with id TMDPDF1. The authors also acknowledge computing time awarded on the Cyclone supercomputer of the High Performance Computing Facility of The Cyprus Institute under project ID p146.
 
\end{acknowledgments}

\bibliography{refs}

\newpage

\appendix

\section{Symmetry Properties of the Quasi-TMD~WF}
\label{sec:symmetry}

To improve the statistical precision of our results, we aim to exploit symmetry properties characterizing the quasi-TMD~WFs. In particular, we seek symmetry relations between correlation functions evaluated at positive and negative values of the staple parameters $b$, $z$ and $L$. By leveraging these relations, we can average over equivalent correlation functions, thereby reducing statistical uncertainties. For this purpose, we examine the behavior of the staple-shaped Wilson line, the pion interpolator and the quasi-TMD~WF correlation function under parity, time reversal and charge conjugation transformations. We perform this investigation in the Euclidean space using the twisted-mass basis; thus, we need first to rotate the operators and interpolators under study from the Minkowski (M) to Euclidean (E) space and from the physical (ph) to the twisted-mass (tw) basis, as follows: 
\begin{eqnarray}
{[{\mathcal{O}_{u\bar{d}}^{\gamma^5\gamma^3}\left(x;b,z,L\right)}]}^{M} &=&
i [{\mathcal{O}_{u\bar{d}}^{\gamma_5\gamma_3}\left(x;b,z,L\right)}]^{E, {\rm ph}} = 
- {\mathcal{O}_{u\bar{d}}^{\gamma_3}}\left(x;b,z,L\right)]^{E, {\rm tw}}, \\
{[{\mathcal{O}_{u\bar{d}}^{\gamma^5\gamma^0}\left(x;b,z,L\right)}]}^{M} &=&
[{\mathcal{O}_{u\bar{d}}^{\gamma_5\gamma_4}\left(x;b,z,L\right)}]^{E, {\rm ph}} = 
i {\mathcal{O}_{u\bar{d}}^{\gamma_4}}\left(x;b,z,L\right)]^{E, {\rm tw}}, \\
{[J_{\pi}(t_0;P^z)]}^{M} &=& {[J_{\pi}(t_0;P^z)]}^{E,{\rm ph}} = {[J_{\pi}(t_0;P^z)]}^{E, {\rm tw}}.
\end{eqnarray}
The definitions of these operators can be found in Section~\ref{sec:setup}.\footnote{Note that the notation used in the main text corresponds to the physical basis in Euclidean space. In this convention, the factor of $i$ in front of the Dirac structure $\gamma_5 \gamma_3$ is omitted, but implicitly understood.}

In Euclidean space, generalized versions of the parity and time reversal transformations defined in each direction $\alpha$ are symmetries of the action. Since we consider twisted-mass fermions, where isospin is broken, it is important to consider transformation symmetries that take into account the flavor content of the operators under study. Thus, we combine the generalized versions of parity and time reversal with a discrete flavor rotation. Below, we define the transformations of the quark and gluon fields under the two generalized symmetries, along with the charge conjugation:

    \begin{align}
    {\rm Parity} \quad & \mathcal{P}_{F\alpha}^{1,2} : \begin{cases} U\left(x_{\alpha},\boldsymbol{x};\alpha\right) \rightarrow U\left(x_{\alpha},-\boldsymbol{x};\alpha\right), \quad U\left(x_{\alpha},\boldsymbol{x};\beta \ne \alpha\right) \rightarrow U^{-1}\left(x_{\alpha},-\boldsymbol{x}-a\hat{\beta};\beta\right), \\ 
    \chi \left(x_{\alpha},\boldsymbol{x}\right) \rightarrow \mathit{i}\gamma_{\alpha}\tau_{1,2}\chi\left(x_{\alpha},-\boldsymbol{x}\right) \\
    \overline{\chi}\left(x_{\alpha},\boldsymbol{x}\right) \rightarrow -\mathit{i}\overline{\chi}\left(x_{\alpha},-\boldsymbol{x}\right)\tau_{1,2}\gamma_{\alpha}
    \end{cases} & \\
    {\rm Time \ Reversal} \quad & \mathcal{T}_{F\alpha}^{1,2} : \begin{cases} U\left(x_{\alpha},\boldsymbol{x};\alpha\right) \rightarrow U^{-1}\left(-x_{\alpha}-a,\boldsymbol{x};\alpha\right), \quad U\left(x_{\alpha},\boldsymbol{x};\beta \ne \alpha\right) \rightarrow U\left(-x_{\alpha},\boldsymbol{x};\beta\right), \\ 
    \chi \left(x_{\alpha},\boldsymbol{x}\right) \rightarrow \mathit{i}\gamma_{\alpha}\gamma_5\tau_{1,2}\chi\left(-x_{\alpha},\boldsymbol{x}\right) \\
    \overline{\chi}\left(x_{\alpha},\boldsymbol{x}\right) \rightarrow -\mathit{i}\overline{\chi}\left(-x_{\alpha},\boldsymbol{x}\right)\tau_{1,2}\gamma_5\gamma_{\alpha}
    \end{cases} & \\
    {\rm Charge \ Conjugation}\quad & \mathcal{C} : \begin{cases}
    U(x) \rightarrow \left(U^{\dagger}(x)\right)^T \\
    \chi (x) \rightarrow C^{-1}\overline{\chi}^T (x) \\
    \overline{\chi} (x) \rightarrow -\chi^T (x) C
    \end{cases} &
    \end{align}
    where $\boldsymbol{x} \equiv x - x_\alpha \hat{\alpha}$, $\chi\left(x_{\alpha},\boldsymbol{x}\right)$ is the fermion field in the twisted mass basis, $U\left(x_{\alpha},\boldsymbol{x};\alpha\right)$ the gauge link in direction $\alpha$ and $\tau_{1,2}$ are the Pauli spin matrices in flavor space.

    We apply the above transformations to the staple-shaped Wilson line, defined in Eq.\eqref{Wilson line}. Without loss of generality, we set the boost and transverse directions of the staple as $\hat{z} = \hat{3}$ and $\hat{n}_{\perp} = \hat{2}$. The resulting relations are listed below. Under parity, we obtain:

\begin{align}
    &\mathcal{W}(t,x_1,x_2,x_3; b, z, L)\xrightarrow{\mathcal{P}_{F_1}^{1,2} } \mathcal{W}(-t,x_1,-x_2,-x_3; -b,-z,-L)&   \\
    &\mathcal{W}(t,x_1,x_2,x_3; b, z, L)\xrightarrow{\mathcal{P}_{F_2}^{1,2} } \mathcal{W}(-t,-x_1,x_2,-x_3; b,-z,-L)&  
    \\
    &\mathcal{W}(t,x_1,x_2,x_3; b, z, L)\xrightarrow{\mathcal{P}_{F_3}^{1,2} } \mathcal{W}(-t,-x_1,-x_2,x_3; -b,z,L)&  
    \\
    &\mathcal{W}(t,x_1,x_2,x_3; b, z, L)\xrightarrow{\mathcal{P}_{F_4}^{1,2} } \mathcal{W}(t,-x_1,-x_2,-x_3; -b,-z,-L).&  
\end{align}

Under generalized time reversal transformations, we obtain:

\begin{align}
    &\mathcal{W}(t,x_1,x_2,x_3; b, z, L)\xrightarrow{\mathcal{T}_{F_1}^{1,2} } \mathcal{W}(t,-x_1,x_2,x_3; b,z,L)&   \\
    &\mathcal{W}(t,x_1,x_2,x_3; b, z, L)\xrightarrow{\mathcal{T}_{F_2}^{1,2} } \mathcal{W}(t,x_1,-x_2,x_3; -b,z,L)&  
    \\
    &\mathcal{W}(t,x_1,x_2,x_3; b, z, L)\xrightarrow{\mathcal{T}_{F_3}^{1,2} } \mathcal{W}(t,x_1,x_2,-x_3; b,-z,-L)&  
    \\
    &\mathcal{W}(t,x_1,x_2,x_3; b, z, L)\xrightarrow{\mathcal{T}_{F_4}^{1,2} } \mathcal{W}(-t,x_1,x_2,x_3; b,z,L).&  
\end{align}

Under charge conjugation transformations, we obtain:
\begin{align}
    \mathcal{W}(t,x_1,x_2,x_3; b, z, L)\xrightarrow{\mathcal{C}} \left(\mathcal{W}(t,x_1,x_2+b,x_3; -b,-z,L)\right)^T. 
\end{align}

These relations, combined with the transformations of the quark fields, allow us to derive the following relations for the bare quasi-TMD~WF correlation function $C_{wf}$ defined in Eq.\eqref{eq:twop}. We focus on the two main leading-twist contributions, coming from $\Gamma_{\phi}=\gamma^5\gamma^0$ and $\Gamma_{\phi}=\gamma^5\gamma^3$ in the Minkowski space, or equivalently $\Gamma_{\phi}=\mathit{i}\gamma_4$ and $\Gamma_{\phi}=-\gamma_3$ in the twisted-mass basis in Euclidean space. Given that the three symmetries affect the flavors of the two quark propagators involved in $C_{wf}$, we include below two flavor indices in our notation.

\begin{align}
    &{C_{wf}}_{du}\left(t,P^z,-\gamma_3;b,z,L\right)\xrightarrow{\mathcal{P}_{F_1}^{1,2}}{C_{wf}}_{ud}\left(-t,-P^z,-\gamma_3;-b,-z,-L\right)&  
    \\
    &{C_{wf}}_{du}\left(t,P^z,-\gamma_3;b,z,L\right)\xrightarrow{\mathcal{P}_{F_2}^{1,2}}{C_{wf}}_{ud}\left(-t,-P^z,-\gamma_3;b,-z,-L\right)&  
    \\
    &{C_{wf}}_{du}\left(t,P^z,-\gamma_3;b,z,L\right)\xrightarrow{\mathcal{P}_{F_3}^{1,2}}-{C_{wf}}_{ud}\left(-t,P^z,-\gamma_3;-b,z,L\right)&  
    \\
    &{C_{wf}}_{du}\left(t,P^z,-\gamma_3;b,z,L\right)\xrightarrow{\mathcal{P}_{F_4}^{1,2}}{C_{wf}}_{ud}\left(t,-P^z,-\gamma_3;-b,-z,-L\right)&  
\end{align}

\begin{align}
    &{C_{wf}}_{du}\left(t,P^z,\mathit{i}\gamma_4;b,z,L\right)\xrightarrow{\mathcal{P}_{F_1}^{1,2}}{C_{wf}}_{ud}\left(-t,-P^z,\mathit{i}\gamma_4;-b,-z,-L\right)&  
    \\
    &{C_{wf}}_{du}\left(t,P^z,\mathit{i}\gamma_4;b,z,L\right)\xrightarrow{\mathcal{P}_{F_2}^{1,2}}{C_{wf}}_{ud}\left(-t,-P^z,\mathit{i}\gamma_4;b,-z,-L\right)&  
    \\
    &{C_{wf}}_{du}\left(t,P^z,\mathit{i}\gamma_4;b,z,L\right)\xrightarrow{\mathcal{P}_{F_3}^{1,2}}{C_{wf}}_{ud}\left(-t,P^z,\mathit{i}\gamma_4;-b,z,L\right)&  
    \\
    &{C_{wf}}_{du}\left(t,P^z,\mathit{i}\gamma_4;b,z,L\right)\xrightarrow{\mathcal{P}_{F_4}^{1,2}}-{C_{wf}}_{ud}\left(t,-P^z,\mathit{i}\gamma_4;-b,-z,-L\right)&  
\end{align}

\begin{align}
    &{C_{wf}}_{du}\left(t,P^z,-\gamma_3;b,z,L\right)\xrightarrow{\mathcal{T}_{F_1}^{1,2}}-{C_{wf}}_{ud}\left(t,P^z,-\gamma_3;b,z,L\right)&  
    \\
    &{C_{wf}}_{du}\left(t,P^z,-\gamma_3;b,z,L\right)\xrightarrow{\mathcal{T}_{F_2}^{1,2}}-{C_{wf}}_{ud}\left(t,P^z,-\gamma_3;-b,z,L\right)&  
    \\
    &{C_{wf}}_{du}\left(t,P^z,-\gamma_3;b,z,L\right)\xrightarrow{\mathcal{T}_{F_3}^{1,2}}{C_{wf}}_{ud}\left(t,-P^z,-\gamma_3;b,-z,-L\right)&  
    \\
    &{C_{wf}}_{du}\left(t,P^z,-\gamma_3;b,z,L\right)\xrightarrow{\mathcal{T}_{F_4}^{1,2}}-{C_{wf}}_{ud}\left(-t,P^z,-\gamma_3;b,z,L\right)&  
\end{align}

\begin{align}
    &{C_{wf}}_{du}\left(t,P^z,\mathit{i}\gamma_4;b,z,L\right)\xrightarrow{\mathcal{T}_{F_1}^{1,2}}-{C_{wf}}_{ud}\left(t,P^z,\mathit{i}\gamma_4;b,z,L\right)&  
    \\
    &{C_{wf}}_{du}\left(t,P^z,\mathit{i}\gamma_4;b,z,L\right)\xrightarrow{\mathcal{T}_{F_2}^{1,2}}-{C_{wf}}_{ud}\left(t,P^z,\mathit{i}\gamma_4;-b,z,L\right)&  
    \\
    &{C_{wf}}_{du}\left(t,P^z,\mathit{i}\gamma_4;b,z,L\right)\xrightarrow{\mathcal{T}_{F_3}^{1,2}}-{C_{wf}}_{ud}\left(t,-P^z,\mathit{i}\gamma_4;b,-z,-L\right)&  
    \\
    &{C_{wf}}_{du}\left(t,P^z,\mathit{i}\gamma_4;b,z,L\right)\xrightarrow{\mathcal{T}_{F_4}^{1,2}}{C_{wf}}_{ud}\left(-t,P^z,\mathit{i}\gamma_4;b,z,L\right)&  
\end{align}

\begin{align}
    &{C_{wf}}_{du}\left(t,P^z,-\gamma_3;b,z,L\right)\xrightarrow{\mathcal{C}}-{C_{wf}}_{ud}\left(t,P^z,-\gamma_3;-b,-z,L\right)&  
    \\
    &{C_{wf}}_{du}\left(t,P^z,\mathit{i}\gamma_4;b,z,L\right)\xrightarrow{\mathcal{C}}-{C_{wf}}_{ud}\left(t,P^z,\mathit{i}\gamma_4;-b,-z,L\right).&  
\end{align}

By combining two such transformations, we obtain the following relations between quasi-TMD WFs that do not involve swapped flavor indices:

\begin{align}
    &{C_{wf}}_{du}\left(t,P^z,-\gamma_3;b,z,L\right)\xrightarrow{\mathcal{P}_{F_1}^{1,2}\cdot\mathcal{P}_{F_2}^{1,2}}{C_{wf}}_{du}\left(t,P^z,-\gamma_3;-b,z,L\right)&  
    \\
    &{C_{wf}}_{du}\left(t,P^z,-\gamma_3;b,z,L\right)\xrightarrow{\mathcal{P}_{F_1}^{1,2}\cdot\mathcal{P}_{F_3}^{1,2}}-{C_{wf}}_{du}\left(t,-P^z,-\gamma_3;b,-z,-L\right)&  
    \\
    &{C_{wf}}_{du}\left(t,P^z,-\gamma_3;b,z,L\right)\xrightarrow{\mathcal{P}_{F_1}^{1,2}\cdot\mathcal{P}_{F_4}^{1,2}}{C_{wf}}_{du}\left(-t,P^z,-\gamma_3;b,z,L\right)& \label{Relation_t_1} 
\end{align}

\begin{align}
    &{C_{wf}}_{du}\left(t,P^z,\mathit{i}\gamma_4;b,z,L\right)\xrightarrow{\mathcal{P}_{F_1}^{1,2}\cdot\mathcal{P}_{F_2}^{1,2}}{C_{wf}}_{du}\left(t,P^z,\mathit{i}\gamma_4;-b,z,L\right)&  
    \\
    &{C_{wf}}_{du}\left(t,P^z,\mathit{i}\gamma_4;b,z,L\right)\xrightarrow{\mathcal{P}_{F_1}^{1,2}\cdot\mathcal{P}_{F_3}^{1,2}}{C_{wf}}_{du}\left(t,-P^z,\mathit{i}\gamma_4;b,-z,-L\right)&  
    \\
    &{C_{wf}}_{du}\left(t,P^z,\mathit{i}\gamma_4;b,z,L\right)\xrightarrow{\mathcal{P}_{F_1}^{1,2}\cdot\mathcal{P}_{F_4}^{1,2}}-{C_{wf}}_{du}\left(-t,P^z,\mathit{i}\gamma_4;b,z,L\right)& \label{Relation_t_2} 
\end{align}

\begin{align}
    &{C_{wf}}_{du}\left(t,P^z,-\gamma_3;b,z,L\right)\xrightarrow{\mathcal{C}\cdot\mathcal{T}_{F_2}^{1,2}}{C_{wf}}_{du}\left(t,P^z,-\gamma_3;b,-z,L\right)&  
    \\
    &{C_{wf}}_{du}\left(t,P^z,\mathit{i}\gamma_4;b,z,L\right)\xrightarrow{\mathcal{C}\cdot\mathcal{T}_{F_2}^{1,2}}{C_{wf}}_{du}\left(t,P^z,\mathit{i}\gamma_4;b,-z,L\right)&   
\end{align}

Additional relations can be derived by considering the complex conjugate of $C_{wf}$. Note that in Euclidean space, the Hermitian conjugate of an operator $\mathcal{O} (t,\vec{x})$ exhibits a sign flip in its time dependence, as can be demonstrated using the Heisenberg picture:
\begin{equation}
  \mathcal{O} (t,\vec{x}) = e^{Ht} \mathcal{O} (0,\vec{x}) e^{-Ht} \Rightarrow \mathcal{O}^\dagger (t,\vec{x}) = e^{-Ht} \mathcal{O}^\dagger (0,\vec{x}) e^{Ht}. 
\end{equation}
The resulting relations are given below:
\begin{eqnarray}
  {C_{wf}^\ast}_{du}\left(t,P^z,-\gamma_3;b,z,L\right) &=& {C_{wf}^\ast}_{ud}\left(-t,-P^z,-\gamma_3;-b,-z,L\right), \\
  {C_{wf}^\ast}_{du}\left(t,P^z,\mathit{i}\gamma_4;b,z,L\right) &=& {C_{wf}^\ast}_{ud}\left(-t,-P^z,\mathit{i}\gamma_4;-b,-z,L\right).
  \end{eqnarray}
  Applying now symmetry transformations to the complex conjugate of the correlation functions, we obtain further relations, e.g.,
\begin{align}
    &{C_{wf}^\ast}_{du}\left(t,P^z,-\gamma_3;b,z,L\right)\xrightarrow{\mathcal{P}_{F_1}^{1,2}} {C_{wf}}_{du}\left(t,P^z,-\gamma_3;b,z,-L\right)&  
    \\
&{C_{wf}^\ast}_{du}\left(t,P^z,\mathit{i}\gamma_4;b,z,L\right)\xrightarrow{\mathcal{P}_{F_1}^{1,2}} {C_{wf}}_{du}\left(t,P^z,\mathit{i}\gamma_4;b,z,-L\right)& 
\end{align}
and
\begin{align}
    &{C_{wf}^\ast}_{du}\left(t,P^z,-\gamma_3;b,z,L\right)\xrightarrow{\mathcal{C} \cdot \mathcal{P}_{F_1}^{1,2} \cdot \mathcal{P}_{F_4}^{1,2}} -{C_{wf}}_{du}\left(t,-P^z,-\gamma_3;b,z,L\right)& \label{Relation_Pz_1} 
    \\
&{C_{wf}^\ast}_{du}\left(t,P^z,\mathit{i}\gamma_4;b,z,L\right)\xrightarrow{\mathcal{C} \cdot \mathcal{P}_{F_1}^{1,2} \cdot \mathcal{P}_{F_4}^{1,2}} {C_{wf}}_{du}\left(t,-P^z,\mathit{i}\gamma_4;b,z,L\right).& \label{Relation_Pz_2}
\end{align}

As concluded from the above relations, the quasi-TMD WFs under study must be symmetric in both $b$ and $z$ for both real and imaginary parts and for both Dirac structures under study. In addition, the real (imaginary) part must be symmetric (antisymmetric) in $L$. Moreover, the real (imaginary) part must be symmetric (antisymmetric) in $P^z$ for the Dirac structure $i\gamma_4$, while the opposite is valid for $-\gamma_3$. 

\section{Fitting Procedure}
\label{sec:fitting}

\begin{figure}[h!]
    \centering
    \subfigure{\includegraphics[width=0.49\linewidth]{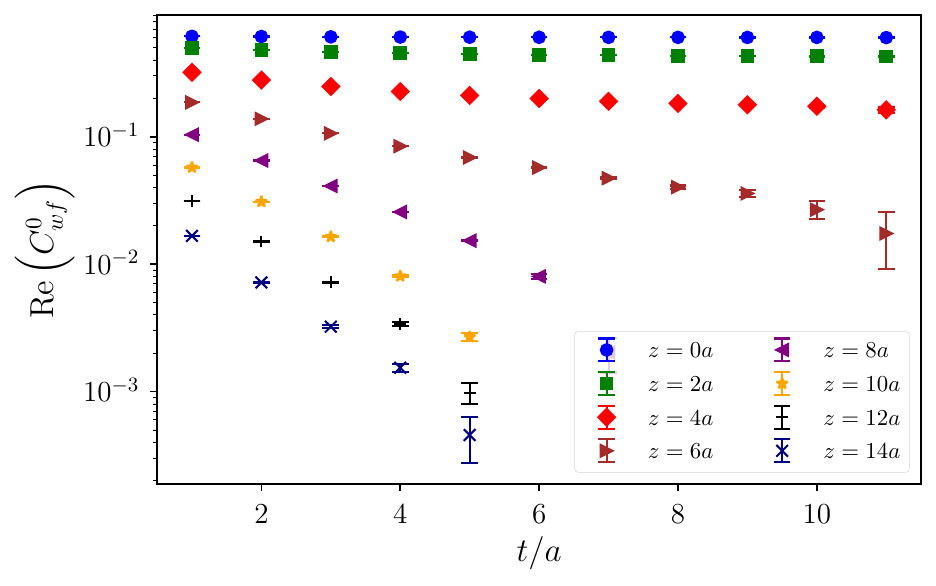}}%
    \subfigure{\includegraphics[width=0.49\linewidth]{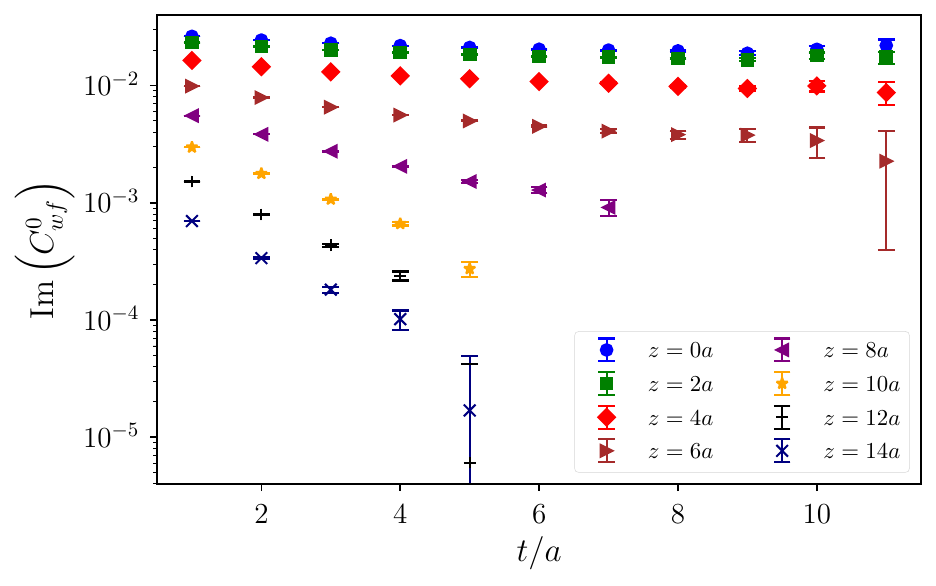}}%
    
    \caption{Dependence of the real(left) and imaginary (right) contributions of the normalized quasi-TMD~WF on the sink time $t$ in logarithmic scale.  Results obtained with $b=1a$ and $L=8a$ on the cA211.53.24 ensemble, valence pion mass $m_{\pi} = 830$~MeV and boost $P^z = 2.22$~GeV, for $\Gamma_\phi = \gamma_5 \gamma_4$.
    }
    \label{fig:excited}
\end{figure}

In this appendix, we outline the fitting procedure used to extract the normalized quasi-TMD~WF, $\psi_{\Gamma_\phi}^{\rm norm}$ (Eq.~\eqref{norm wavefunction}), and the normalized meson form factor, $F^{\rm norm}_{\Gamma}$ (Eq.~\eqref{Fnorm}). Our analysis is based on the spectral decomposition of the normalized correlation functions $C^0_{wf}$ and $C^0_{4pt}$, provided in Eqs.~(\ref{eq:Cwf0}--\ref{eq:C4pt0}). The fits include either only the ground-state contribution (one-state fit) or both the ground and first excited states (two-state fit). The explicit fitting functions used for the Dirac structures relevant to our analysis are as follows:
\begin{eqnarray} 
\text{One-state:} && \quad C_{wf}^0\left(t,P^z,\Gamma_{\phi};b,z,L\right) = \tilde{\psi}^{\rm norm}_{\Gamma_{\phi}}(b,z,L;P^z), \\
\text{Two-state:} && \quad C_{wf}^0\left(t,P^z,\Gamma_{\phi};b,z,L\right) = \tilde{\psi}^{\rm norm}_{\Gamma_{\phi}}(b,z,L;P^z) \times \frac{1 + c_1^{\Gamma_\phi}(b,z,L;P^z) \, e^{-\Delta E^{\prime}(P^z) t}}{1 + c_1^{\gamma_5 \gamma_4}(0,0,0;P^z) \, e^{-\Delta E(P^z) t}}, \label{two-state Cwf} \\
\text{One-state:} && \quad C_{4pt}^0\left(t,t_s,P^z,\Gamma;b\right) = N_\Gamma \, F_\Gamma^{\rm norm}(b,P^z), \\
\text{Two-state:} && \quad C_{4pt}^0\left(t,t_s,P^z,\Gamma;b\right) = \frac{F_\Gamma^{(0,0)} (b;P^z)}{\tilde{\psi}^{(0)}_{\Gamma_{\phi}} (0,0,0;-P^z) \,\tilde{\psi}^{(0)}_{\Gamma_{\phi}} (0,0,0;P^z)} \times \nonumber \\
    && \frac{1+d_{0,1}^{\Gamma}(b;P^z) e^{- \Delta E (P^z) t} + d_{1,0}^{\Gamma}(b;P^z) e^{- \Delta E (P^z) (t_s - t)} + d_{1,1}^{\Gamma}(b;P^z) e^{- \Delta E (P^z) t_s} + \ldots}{\left[1 + c_1^{\Gamma_\phi}(0,0,0;-P^z) \, e^{-\Delta E (P^z) t_s/2} + \ldots\right] \times \left[1 + c_1^{\Gamma_\phi}(0,0,0;P^z) \, e^{-\Delta E (P^z) t_s/2} + \ldots\right]} = \nonumber \\ && = N_\Gamma \, F_\Gamma^{\rm norm}(b,P^z) \times \frac{1 + d_{0,1}^{\Gamma}(b;P^z)\left[e^{-\Delta E(P^z) t} + e^{-\Delta E(P^z) (t_s - t)}\right]}{\left[1 + c_1^{\gamma_5 \gamma_4}(0,0,0;P^z) \, e^{-\Delta E(P^z) t_s/2}\right]^2}, \label{2-state 4pt}
\end{eqnarray}
where $c_1^{\Gamma_\phi}(b,z,L;P^z)$ is a function of $Z^{(0)}$, $Z^{(n)}$, $E^{(0)}$, $E^{(n)}$, $\tilde{\psi}^{(0)}_{\Gamma_{\phi}}$, $\tilde{\psi}^{(n)}_{\Gamma_{\phi}}$, and $d_{n,m}^{\Gamma} (b;P^z)$ is a function of $Z^{(0)}$, $Z^{(n)}$, $Z^{(m)}$, $E^{(0)}$, $E^{(n)}$, $E^{(m)}$, $F_\Gamma^{(0,0)}$, $F_\Gamma^{(n,m)}$; Note that $d_{0,1}^{\Gamma} (b;P^z) = d_{1,0}^{\Gamma} (b;P^z)$ and we drop  $d_{1,1}^{\Gamma} (b;P^z)$ since it is a higher order term. Also, $c_1^{\gamma_5 \gamma_4}(0,0,0;P^z) = c_1^{\gamma_5 \gamma_4}(0,0,0;-P^z)$ and $c_1^{\gamma_5 \gamma_3}(0,0,0;P^z) = -c_1^{\gamma_5 \gamma_3}(0,0,0;-P^z)$; $\Delta E(P^z) \equiv E^{(1)}(P^z) - E^{(0)}(P^z)$ is the energy gap between the first excited and ground states obtained at $z=0$. In Eq.~\eqref{two-state Cwf}, we allow  the excited state to be different for the numerator and denominator, namely we take  $\Delta E'(P^z) \equiv E'^{(n)}(P^z) - E^{(0)}(P^z) \neq \Delta E(P^z)$ for each $z$. As shown in Fig.~\ref{fig:excited}, the contamination that may depend on overlap terms varies for different $z$.

The Dirac structures used are $\Gamma_\phi = \gamma_5 \gamma_3, \gamma_5 \gamma_4$ and $\Gamma = \gamma_5 \gamma_\perp, \gamma_\perp, \gamma_5 \gamma_j, \gamma_j$ (see main text for details).

The two-state fits are expected to provide a more accurate determination of the ground-state contribution, as they allow for a wider fitting range and incorporate additional data points with better signal-to-noise ratios, particularly near the source. To keep the functional form manageable, we omit the term proportional to $d_{1,1}^\Gamma(b;P^z)$ in the second line from Eq.~\eqref{2-state 4pt} for the two-state fit.

The one-state fit involves a single parameter, while the two-state fit includes five for the the normalized quasi-TMD~WF and four for the normalized meson form factor. Two of these, $c_1^{\Gamma_\phi}(0,0,0;P^z)$ and $\Delta E(P^z)$, are determined independently by fitting the logarithm of the ratio of correlation functions at consecutive time steps (in lattice units), assuming large $t$:

\begin{align}
\lim_{t \gg 0} \log \left[\frac{C_{wf}\left(t,P^z,\Gamma_{\phi};0,0,0\right)}{C_{wf}\left(t+1,P^z,\Gamma_{\phi};0,0,0\right)}\right] \approx E^{(0)}(P^z) + \log\left[\frac{1 + c_1(0,0,0;P^z) \, e^{-\Delta E(P^z) t}}{1 + c_1(0,0,0;P^z) \, e^{-\Delta E(P^z) (t+1)}}\right]. \label{eq:ratio}
\end{align}

The extracted values of $c_1^{\Gamma_\phi}(0,0,0;P^z)$ and $\Delta E(P^z)$ are then fixed in the two-state fits of Eqs.~(\ref{two-state Cwf},~\ref{2-state 4pt}). Using Eq.~\ref{two-state Cwf}, we perform a joint fit between the real and imaginary parts of $C_{wf}^0$, while for $C_{4pt}^0$ we perform a joint fit over the different values of $t_s$ using Eq.~\ref{2-state 4pt}.

We perform both one- and two-state fits using a least-squares method that minimizes the reduced correlated chi-squared, $\chi^2/{\rm d.o.f.}$, using the full covariance matrix. This approach accounts for correlations among data points, leading to more reliable uncertainty estimates compared to fits using only uncorrelated standard deviations.

The fitting range $t \in [t_{\rm low}, t_{\rm high}]$ for $C^0_{wf}$ is determined as follows:
\begin{itemize}
    \item $t_{\rm high}$ is chosen as the time slice at which the signal-to-noise ratio drops below one third, indicating unreliable data due to noise dominance.
    \item $t_{\rm low}$ is varied within $[2a,5a]$ for the two-state fits, and within $[6a, t_{\rm high}]$ for the one-state fits. For each $t_{\rm low}$, we extract the ground-state contribution and select the two-state result at the value where convergence with the one-state fit is observed, indicating suppression of excited-state contamination.
\end{itemize}

For the $C^0_{4pt}$ correlator, the fitting range is chosen as $t \in [2a, t_s - 2a]$ for both the two-state fits and one-state fits, for each value of $t_s$. 

In all selected cases, we confirm that the reduced correlated chi-squared value is close to unity, ensuring the quality and consistency of our fits.
 
\end{document}